\documentclass[12pt,preprint]{aastex}

\def\feka{Fe K$\alpha$}
\def\chandra{{\it Chandra}}
\def\xmm{{\it XMM-Newton}}
\def\suzaku{{\it Suzaku}}
\def\swift{{\it Swift}}
\def\asca{{\it ASCA}}

\def\rxte{{\it RXTE}}
\def\sax{{\it BeppoSAX}}

\def\integral{{\it Integral}}

\def\fermi{{\it Fermi}}

\def\lum{erg s$^{-1}$}
\def\flux{erg cm$^{-2}$ s$^{-1}$}
\def\nh{cm$^{-2}$}
\def\arcsec{$^{\prime\prime}$}
\def\deg{$^{\circ}$}

\def\ltsima{$\; \buildrel < \over \sim \;$}
\def\simlt{\lower.5ex\hbox{\ltsima}} 
\def\gtsima{$\; \buildrel > \over \sim \;$}
\def\simgt{\lower.5ex\hbox{\gtsima}} 

\def\3c{3C~390.3}

\input{psfig.sty}

\begin{document}

\title{Structure of the Accretion Flow in Broad-Line Radio Galaxies: 
The Case of 3C~390.3}

\author{R. M. Sambruna}
\affil{NASA/GSFC, Code 661, Greenbelt, MD 20771}

\author{J. N. Reeves}
\affil{NASA/GSFC, Code 662, Greenbelt, MD 20771, and
Astrophysics Group, School of Physical \& Geographical Sciences, Keele
University, Keele, Staffordshire ST5 5BG, UK}

\author{V. Braito}
\affil{Department of Physics and Astronomy, Johns Hopkins University,
Baltimore, MD 21218, and 
University of Leicester, Department of Physics \& Astronomy,
University Road, Leicester LE1 7RH, UK}

\author{K. T. Lewis}
\affil{NASA/GSFC, Code 662, Greenbelt, MD 20771 and 
Dickinson College, Dept. of Physics and Astronomy, Carlisle, PA 17013}

\author{M. Eracleous}
\affil{Department of Astronomy \& Astrophysics, The Pennsylvania State
University, 525 Davey Lab, University Park, PA
16802, and Center for Gravitational Wave Physics, The Pennsylvania
  State University, University Park, PA 16802}

\author{M. Gliozzi}
\affil{George Mason University, Department of Physics and Astronomy,
  4400 University Dr., Fairfax, VA, 22030} 

\author{F. Tavecchio}
\affil{INAF, Osserv. Astron. di Brera, via Bianchi 46, I-23807 Merate, Italy}

\author{D.R. Ballantyne}
\affil{Center for Relativistic Astrophysics, School of Physics,
Georgia Institute of Technology, 837 State Street, Atlanta, GA 30332}

\author{P. M. Ogle}
\affil{Spitzer Science Center, California Institute of Technology,
Mail Code 220-6, Pasadena, CA 91125}

\author{A.J. Barth}
\affil{Department of Physics and Astronomy, 4129 Frederick Reines Hall,
University of California, Irvine, CA 92697-4575}

\author{J. Tueller} 
\affil{NASA/GSFC, Code 661, Greenbelt, MD 20771}

%
%
%

\begin{abstract}

We present \xmm\ and \suzaku\ observations of the Broad-Line Radio
Galaxy (BLRG) \3c\ acquired in October 2004 and December 2006,
respectively. An archival \swift\ BAT spectrum from the 9 month survey
is also analyzed, as well as an optical spectrum simultaneous to
\xmm. At soft X-rays, no absorption features are detected in the RGS
spectrum of \3c; a narrow emission line is found at 0.564~keV, most
likely originating in the Narrow Line Region. Both the EPIC and XIS
datasets confirm the presence of an \feka\ emission line at 6.4~keV
with EW=40~eV.  The \feka\ line has a width FWHM $\sim$ 8,800 km/s,
consistent within a factor two with the width of the double-peaked
H$\alpha$ line, suggesting an origin from the Broad Line Region. The
data show for the first time a weak, broad bump extending from 5 to
7~keV. When fitted with a Gaussian, its centroid energy is 6.6~keV in
the source's rest-frame with FWHM of 43,000 km/s and EW of 50~eV; its
most likely interpretation is emission from He-like Fe (Fe~XXV),
suggesting the presence of an ionized medium in the inner regions of
\3c. The broad-band 0.5--100~keV continuum is well described by a
single power law with photon index $\Gamma=1.6$ and cutoff energy
157~keV, plus cold reflection with strength $R=0.5$. In addition,
ionized reflection is required to account for the 6.6~keV bump in the
broad-band continuum, yielding an ionization parameter $\xi \sim 2700$
ergs cm s$^{-1}$; the inner radius of the ionized reflector is
constrained to be larger than 20$r_G$, although this result depends on
the assumed emissivity profile of the disk. If true, we argue that the
lack of broad Fe~K emission from within 20$r_G$ indicates that the
innermost regions of the disk in \3c\ are obscured and/or poorly
illuminated. While the SED of \3c\ is generally dominated by
accretion-related continuum, during accretion low states the jet can
significantly contribute in the optical to X-ray bands via synchrotron
self-Compton emission.  The Compton component is expected to extend to
and peak at GeV gamma-rays where it will be detected with the
\fermi\ Gamma-Ray Space Telescope during its first few years of
operation.

\end{abstract}

{\sl Subject Headings:}{Galaxies: active --- galaxies: radio --
galaxies: individual --- X-rays: galaxies}

\section{Introduction}

According to a widely accepted view Active Galactic Nuclei (AGN),
including radio-loud sources, are ultimately powered by accretion onto
a central supermassive black hole (e.g., Blandford 1985). A
relativistic jet originates in the innermost regions and transports
energy and momentum to the more distant lobes of the radio galaxy. The
jet angle -- defined as the angle between the jet axis and the line of
sight -- increases from blazars to Broad- and Narrow-Line Radio
Galaxies, in which the direct view to the center is obscured by thick
matter surrounding the nucleus. As the jet angle increases, the
importance of its emission decreases, due to beaming effects.

Previous X-ray observations of BLRGs established that these sources
exhibit Seyfert-like spectra with subtle but significant
differences. Specifically, the X-ray continua of BLRGs are flatter
($\langle\Gamma\rangle \approx 1.7$; Zdziarski \& Grandi 2001), and
their reflection features weaker than their radio-quiet cousins (e.g.,
Sambruna, Eracleous, \& Mushotzky 2002; Ballantyne 2007, and
references therein). To account for these properties it was suggested
that the reflection features from a standard, cold disk similar to
Seyferts are weak in BLRGs because they are diluted by the non-thermal
beamed jet emission (Grandi, Urry, \& Maraschi 2002). Another
possibility is that the reprocessing medium in radio-loud objects subtends a
smaller solid angle to the central X-ray source, as in ion
torus/advection dominated accretion flow models (Rees et al. 1982), as
advocated by Eracleous, Sambruna, \& Mushotzky (2000). Alternatively,
BLRGs could have more highly ionized accretion disks than Seyfert 1s,
as a result of higher accretion rates (Ballantyne, Ross, \& Fabian
2002). This model successfully fitted the \xmm\ spectrum of 3C~120
(Ballantyne, Fabian, \& Iwasawa 2004). However, fitting the same
dataset Ogle et al. (2002) found no evidence for broad Fe-K emission
line, and concluded based on the line width and equivalent width that
the \feka\ line originates from the optical BLR.  The various
scenarios discussed above are not mutually exclusive, however, and it
is possible that jet dilution could affect the strength of the
spectroscopic signature of an ionized disk, for example. However, from
the analysis of \sax\ observations of BLRGs, Grandi \& Palumbo (2007)
concluded that if present, the jet contributes minimally to the
2--10~keV emission but should dominate at GeV gamma-rays. 

The key observational features required to disentangle the various
possibilities are broad-band coverage and sensitivity at medium-hard
X-rays, where the reflection features and the jet non-thermal emission
mix.  Previous attempts at acquiring broad-band coverage at these
energies by coordinating \xmm\ and \rxte\ had modest success due to
the different scheduling constraints of the two missions, while
\sax\ had limited sensitivity in the Fe~K line region. It was not
until the advent of \suzaku\ that the basic requisites of simultaneous
broadband coverage and larger collecting area in 6--7~keV was
achieved. Thus, we began a program of observations of a number of
``classical'' (i.e., bright) BLRGs with \suzaku\ (\3c, 3C~382, 3C~445,
3C~111); two additional sources, 3C~120 and 4C~+74.26, are already in
the \suzaku\ archive (Kataoka et al. 2007; Larsson et al. 2008).

Our program consists of two complementary goals: 1) determine if jet
dilution is responsible for the weakness of the reflection features in
BLRGs; and 2) study the structure of the accretion flow in radio-loud
AGN, to understand which conditions favor the formation of powerful,
relativistic jets. The source whose \suzaku\ observations are
presented here, \3c, is an optimal candidate for goal 2. As described
in \S~2 and shown in \S~7.3, the jet does not make a significant
contribution to the nuclear X-ray emission of \3c, and as such this
BLRG qualifies as an optimal laboratory for investigating the
accretion flow structure in radio-loud AGN.

In this paper we also present the results of our \xmm-optical
spectroscopy campaign of \3c\ performed in October 2004. An optical
spectrum was acquired at Keck Observatory simultaneously to the
\xmm\ EPIC observations in order to enable us to compare the profiles
of the \feka\ and double-peaked Balmer lines, testing the idea that
the former originates in the disk/Broad Line Region (BLR). While this
paper focuses on a spectroscopic study, a detailed discussion of the
X-ray timing properties of \3c\ based on the various datasets can be
found in a companion paper by Gliozzi et al. (2009).

The paper is organized as follows. After describing the source
properties and previous observations in \S~2, in \S~3 the data
reduction of the new observations is presented. We describe the
spectral fits to the various datasets in \S~4. In \S~5 we discuss the
multiwavelength SED.  Results are summarized in \S~6 and discussed in
\S~7, with conclusions following in \S~8. Throughout this paper, a
concordance cosmology with H$_0=71$ km s$^{-1}$ Mpc$^{-1}$,
$\Omega_{\Lambda}$=0.73, and $\Omega_m$=0.27 (Spergel et al. 2003) is
adopted. The energy spectral index, $\alpha$, is defined such that
$F_{\nu} \propto \nu^{-\alpha}$. The photon index is $\Gamma=\alpha+1$.

\section{Properties of 3C~390.3} 

The Broad-Line Radio Galaxy \3c, located at $z$=0.056, is well studied
at all wavelengths from radio to X-rays. In the radio it is a
classical double-lobed FRII radio galaxy (Pearson \& Readhead 1988)
and the core exhibits superluminal motion (Alef et al. 1996). From the
apparent velocity and the core dominance, a jet inclination angle with
respect to the line of sight of 30\deg\ $<\theta<$ 35\deg\ was
inferred (Giovannini et al. 2001). 


We can assess the jet contribution to the nuclear emission by
calculating the expected Doppler factor from the jet inclination
angle, 30\deg\ $<\theta<$ 35\deg, and jet velocity, $\beta \sim
0.96-0.99$ (Giovannini et al. 2001). This yields Lorentz factors
$\Gamma_L$ in the range 3.6--7.1 and thus Doppler factors
$\delta=0.6-1.7$. While the range of Doppler factors formally does
not exclude jet amplification, we show in \S~7.3 by direct modeling of
the Spectral Energy Distribution that the beamed jet emission does not
make a significant contribution to the optical-to-X-ray emission from
the nucleus unless the accretion-flow emission is in a very low state,
which was not the case for the observations presented here.

In the optical, \3c\ is known for exhibiting double-peaked optical
emission lines (Eracleous \& Halpern 1994). By fitting the line
profiles with a disk model, an inclination angle between the disk axis
and the line of sight can be derived; for \3c\ this gave
$i=(26^{+4}_{-2})$\deg\ (Eracleous \& Halpern 1994), very close to the
range derived from the radio. Our optical spectrum simultaneous to
\xmm\ provides similar constraints (\S~4.5). 

The source is also well studied in the X-ray band (Grandi et al. 1999
and references therein). At these wavelengths it is variable, showing
large-amplitude variations at both soft and hard X-rays on timescales
of weeks to months (Leighly \& O'Brien 1997; Gliozzi, Sambruna, \&
Eracleous 2003). It was observed with all previous X-ray
observatories, including \asca, which provided the first high-quality
X-ray spectrum (Eracleous, Halpern, \& Livio 1996). The \asca\ 40~ks
spectrum showed a rather hard X-ray continuum with photon index
$\Gamma=1.7$, and a resolved \feka\ emission line at 6~keV with an
Equivalent Width EW$\sim 170$ eV and a Gaussian profile consistent
with an origin from the outer ($r> 250 r_g$) regions of the disk.

A \sax\ observation of \3c\ confirmed the presence of the \feka\ line
and showed a strong reflection bump, $R \sim 1$ (Grandi et
al. 1999). \3c\ was recently detected above 10~keV with the BAT
experiment onboard \swift\ (Tueller et al. 2008; see \S~3.4) and with
\integral\ (Bird et al. 2007). The BAT data are used in this paper
and fitted jointly with \suzaku. 

The Galactic column density in the direction to \3c\ is $3.5 \times
10^{20}$ \nh\ (Kalberla et al. 2005). Previous X-ray observations
detected excess amount of cold absorption above this value (Eracleous
et al. 1996; Sambruna, Eracleous, \& Mushotzky 1999). Comparing
measurements obtained at various epochs, Grandi et al. (1999) showed
that the column density N$_H$ in \3c\ varies on timescale of years,
but is not correlated to changes of the X-ray continuum flux. There
was a claim for the presence of warm absorption in the form of an
absorption edge at 0.65~keV from a reanalysis of the \asca\ data
(Sambruna et al. 1999).

\section{New Observations}

The log of the X-ray observations conducted in pointed mode is
reported in Table~1. 

\subsection{\xmm} 

We observed \3c\ with \xmm\ on 2004 October 8--9 and 17 for 50~ks and
20~ks, respectively.  All of the EPIC cameras (Str\"uder et al. 2001;
Turner et al. 2001) were operated in small window mode to prevent
photon pile-up, and with medium filters, due to the presence of bright
nearby sources in the field of view.  The recorded events were
screened to remove known hot pixels and other data flagged as bad;
only data with {\tt FLAG=0} were used.  The data were processed using
the latest CCD gain values. For the temporal and spectral analyses,
events corresponding to pattern 0--12 (singles, doubles, triples, and
quadruples) in the MOS cameras and 0--4 (singles and doubles only,
since the pn pixels are larger) in the pn camera were accepted. Arf
and rmf files were created with the \xmm\ Science Analysis Software
\verb+SAS+ v.7.0. Examination of the full--field light curves
revealed the presence of one period of background flaring during the
first observation.  These events were excluded, reducing the effective
exposure time to 33~ks on October 8.  The net count rates are listed
in Table~1 together with the effective exposures after data screening. 

The source light curves and spectra were extracted from circular
regions of radius 35\arcsec\ centered on the source. The background
spectra and light curves were extracted from source-free circular
regions on the same chip as the source and with extraction radius of
35\arcsec\ for the EPIC pn camera, and on a different chip with
extraction radius of 70\arcsec\ for the EPIC MOS cameras.  There are
no signs of pile-up in the pn or MOS cameras according to the {\tt
SAS} task {\tt epatplot}. The pn and MOS data were re-binned in
order to contain 100 and 50 counts per channel, respectively, and
fitted jointly. There is no flux variability within the single EPIC
exposures; details of the timing analysis are given in a companion
paper (Gliozzi et al. 2009). 

The Reflection Grating Spectrometer (RGS) data for the two
observations were reduced and analysed separately and then
compared. The data were reduced using the standard SAS task
\verb+rgsproc+ and the most recent calibration files. The total
exposure times are $\sim$ 69~ks for the observation of October 8 and
$\sim$ 49~ks on October 17 for both RGS1 and RGS2.  The RGS1 and RGS2
spectra were binned at the resolution of the instrument ($\Delta
\lambda\sim 0.1\AA$).  Thanks to the long exposure time and the
brightness of the source we have high signal-to-noise ratio in both of
the RGS exposures: a total of $\sim 82,000$ and $49,000$ net counts
were collected in the two observations, respectively (Table~1).

\subsection{\suzaku\ XIS} 

\suzaku\ observed \3c\ on December 14--16, 2007 for a total exposure
time of 100~ks. We used the cleaned event files obtained from version
2 of the \suzaku\ pipeline processing.  Standard screening criteria
were used, namely, only events outside the South Atlantic Anomaly
(SAA) as well as with an Earth elevation angle (ELV) $ >
5\ensuremath{{}^{\circ }}$ were retained, and Earth day-time elevation
angles (DYE\_ELV) $ > 20\ensuremath {{}^{\circ}}$. Futhermore, data
within 256~s of the SAA passage were excluded and a cut-off rigidity
of $ >6 \,\mathrm{GV}$ was adopted. The net exposure time after
screening was 85~ks. The net source count rates are listed in Table~1.

A serendipitous source was detected in the XIS field of view with a
very soft X-ray spectrum (see Appendix). This source is likely to be a
transient, as it is not present in any archival imaging X-ray
observation of \3c.

The XIS spectra of \3c\ were extracted from a circular region of
2.9$'$ radius centered on the source.  Background spectra were
extracted from four circular regions offset from the main target and
avoiding the serendipitous source and the calibration sources.  The
combined area of these four background regions is twice the area of
the main target region. The XIS response (rmf) and ancillary response
(arf) files were produced, using the latest calibration files
available, with the \textit{ftools} tasks \textit{xisrmfgen} and
\textit{xissimarfgen}, respectively.  The source spectra from the FI
CCDs were summed, and fitted jointly with the BI (the XIS1) spectrum.
The net XIS source spectra were binned with a minimum of 100 counts
per bin.

\subsection{\suzaku\ HXD} 

For the HXD-PIN data reduction and analysis we followed the latest
\suzaku\ data reduction guide (the ABC guide Version
2)\footnote[1]{http://heasarc.gsfc.nasa.gov/docs/suzaku/analysis/abc/}.
We used the rev2 data, which include all four cluster units, and the
best background available, which account for the instrumental
background (Kokubun et
al. 2007)\footnote[2]{ftp://legacy.gsfc.nasa.gov/suzaku/doc/hxd/suzakumemo-2008-03.pdf}.
The source and background spectra were extracted within the common
good time interval and the source spectrum was corrected for the
detector deadtime.  The net exposure time after screening was 90~ks.

The contribution of the diffuse cosmic X-ray background counts was
simulated using the spectral form of Boldt (1987), assuming the
response matrix for diffuse emission, and then added to the
instrumental one.  At the time of this analysis (September 2008) two
instrumental background files are available; background A or ``quick''
background and background D or ``tuned'' background. We adopted the
latter which is the latest release and which suffers lower systematic
uncertainties of about 1.3\%, corresponding to about half uncertainty
of the first release of the NXB.  With this choice of background,
\3c\ is detected up to 70~keV at a level of 23\% above the
background. The count rate in 10--30~keV is 0.153 $\pm$ 0.003 c/s. The
HXD-pin spectrum was binned in order to have a signal-to-noise ratio
greater then 10 in each bin, and the latest response file released by
the instrumental team was used. Because of its extremely soft X-ray
spectrum (see Appendix), the contribution of the XIS transient to the
HXD flux is negligible.

\subsection{\swift\ BAT Observations} 

The BAT spectrum was derived from the 9-months survey archive. The
data reduction and extraction procedure of the 8-channel spectrum is
described in Tueller et al. (2008). To fit the BAT spectrum, we used
the latest calibration response \verb+diagonal.rsp+ and background
files as of June 2008. The source was detected in 15--100~keV with a
count rate (1.64 $\pm$ 0.07) $\times 10^{-3}$ c/s. Inspection of the
long-term on-line BAT light curve of \3c\ shows that there is no
variability on timescales of weeks or longer.

\subsection{Optical Spectroscopy} 

An optical spectrum of \3c\ was obtained on the night of 2004 October
9 UT during the first \xmm\ observation at the Keck-II telescope using
the Echellette Spectrograph and Imager (ESI; Sheinis et al.\ 2002).
The galaxy was observed in ESI echelle mode, which covers 3800--11,000
\AA\ over 10 echelle orders.  The ESI exposure began at 5:26 UT with a
duration of 600~s.  A 0.75\arcsec-wide slit was used, giving an
instrumental dispersion of $\sigma_i \approx 22$ km s$^{-1}$.  The
slit was oriented at the parallactic angle and the airmass was 2.05.

The data were reduced following standard techniques and the spectrum
was extracted with an extraction width of 1\arcsec.  Wavelength
calibration was done using exposures of internal HgNe, Xe, and CuAr
comparison lamps.  Flux calibration and correction for telluric
absorption bands were done using an exposure of the standard star
BD+284211, which was observed in evening twilight.  Due to scattered
clouds during the observation, as well as slit losses, the absolute
flux scale of the spectrum is subject to substantial uncertainty.
After flux calibration, the individual echelle orders were combined
into a single spectrum with a uniform linear dispersion of 0.25 \AA\
pixel$^{-1}$.

\section{Spectral Analysis} 

All spectral fits to the X-ray data were performed using \verb+XSPEC+
v.11.3.2ag, except for the fits with ionized reflection for which
v.12.4.0 was used. The significance of adding free parameters to the
model was evaluated with the F-test, with associated probability
P$_F$. All uncertainties quoted are 90\% for one parameter of interest
unless otherwise noted.

\subsection{The RGS spectra} 

For each epoch, the RGS1 and RGS2 datasets were fitted jointly with a
single absorbed power law model, keeping the absorption fixed to the
Galactic value, in the energy range 0.35--1.8~keV. This gave
$\chi^2$=559/458 and 548/458 for observation A and B,
respectively. Figure~1a-b shows the RGS spectra and the residuals of
the power law and Galactic absorption model, stretched out in two
panels to show the details. The best-fit parameters are: $\Gamma=1.79
\pm 0.02$ for observation A and $\Gamma=1.77 \pm 0.02 $ for
observation B; the observed fluxes are $F_{\rm{0.3-2~keV}}=2.5\times
10^{-11}$ erg cm$^{-2}$ s$^{-1}$ and $F_{\rm{0.3-2~keV}}=2.1\times
10^{-11}$ erg cm$^{-2}$ s$^{-1}$, respectively.

Allowing the absorption to vary we found that the fit improves by
$\Delta \chi^2\sim 36$ and 50 in the two exposures yielding $N_H \sim
6\times 10^{20}$ cm$^{-2}$, in slight excess over the Galactic value.
Fixing the local absorption to the Galactic value and adding a
rest-frame absorber, we find an intrinsic column density of $N^z_{\rm
  H}=(1.0 \pm 0.6) \times 10^{20}$ cm$^{-2}$ and photon index is now $
\Gamma=1.87 \pm 0.05$ for the October 8--9 observation, and $N^z_{\rm
  H}=(2.0 \pm 0.7) \times 10^{20}$ cm$^{-2}$ and $\Gamma=1.93 \pm
0.06$ for the October 17 observation.

Inspection of Figure~1 shows that there are no prominent spectral
features in the RGS data, except for an emission line in the RGS1
datasets of both epochs at $\sim$23.5 \AA\ (0.56~keV). We fitted the
RGS1 from both epochs A and B by adding a Gaussian line to the single
power law model; the RGS2 data were not included in this fit because
the RGS2 CCD chip corresponding to the OVII triplet wavelength range
malfunctioned shortly after launch in 1999 (Pollock 2008). The joint
fit provides a rest-frame energy and flux for the Gaussian of
E$_l=0.564\pm$0.002~keV and F$_l=1.9\pm 0.5 \times 10^{-4}$ ph
cm$^{-2}$ s$^{-1}$ having fixed the line width to its best-fit value,
$\sigma_l=5$~eV; the line EW is $\sim 5$~eV. The feature is
significant, with $\Delta\chi^2=54$ for 2 additional dofs
corresponding to P$_F >$99\%. It can be identified with the forbidden
OVII line from the distant Narrow Line Regions, as observed in
Seyferts (Guainazzi et al. 2005). Although the line is highly
statistically significant, it coincides with the observed wavelength
of the neutral OI edge near 23\AA\ (at z=0), which may add some
systematic uncertainties to the measured value.

In short, our analysis of the high signal-to-noise RGS spectrum of
\3c\ shows that there is no evidence for discrete features in
absorption in the energy range 0.35--1.8~keV. The data can be
adequately fitted with a single power law model with column density
slightly in excess to the Galactic value. An emission line is found at
0.564~keV, and its proposed interpretation is [OVII] emission from the
distant Narrow Line Region.

\subsection{The 0.5--10~keV continuum} 

We first describe spectral fits to the \xmm\ EPIC data focusing on the
continuum in the energy range 0.5--10~keV.  Previous spectral studies
demonstrated that the broadband X-ray continuum of \3c\ cannot be
described by a simple power law, indicating the presence of variable
cold absorption, an iron K$\alpha$ line, and the onset of cold
reflection (see \S~2). Figure~2a shows the residuals of the EPIC data
fitted to a single power law model (only the pn data for both
observations are shown for clarity). Indeed, a single power law
absorbed by Galactic $N_{\rm H}$ provides a poor fit to the EPIC
spectrum of \3c, with $\chi^2/{\rm d.o.f.}=1590.1/1117$ and
$\chi^2$/d.o.f.=956.9/758 for observations A and B,
respectively. Line-like residuals in the 5.5--7~keV energy range are
present. A satisfactory fit is obtained for both \xmm\ observations
when a model including two power laws and a Gaussian line is used
instead. The second power law is highly significant (P$_F \sim$ 99\%)
and parameterizes the onset of the reflection component above 8~keV,
better visible in the \suzaku\ data (Figure~2b). When left free to
vary, the absorbing column density is consistent with Galactic within
large errors.

During observation A, the observed soft and hard X-ray fluxes are,
respectively, $F_{\rm 0.5-2~keV}=2.0\times10^{-11}$\flux\ and $F_{\rm
  2-10~keV}=3.9\times10^{-11}$\flux, and the corresponding
luminosities $L_{\rm 0.5-2~keV}=2.0\times10^{44}{\rm~erg~s^{-1}}$ and
$L_{\rm 2-10~keV}=2.9\times10^{44}{\rm~erg~s^{-1}}$.  A modest
decrease of $\sim$ 12--13\% in the flux and luminosity values is
observed during observation B with no accompanying spectral changes
(Gliozzi et al. 2009). Thus, the EPIC pn spectra from observation
A and B were fitted jointly; the results are reported in Table~2a. 

In the joint EPIC fits if left free to vary the column density
converges to a value in excess of Galactic. Thus we added a second
absorber to the model, N$_H^z$, leaving the column free to vary and
fixing the absorber's redshift to be equal to that of the
source. Table~2a shows that the excess column density, if intrinsic,
amounts to $\Delta N_H \sim 2 \times 10^{20}$ \nh\ at 90\% confidence.

Next, we examined the \suzaku\ XIS data.  Spectral fits were performed
in the energy ranges 0.5--10~keV where the calibration is best and the
background negligible. We first fitted the data jointly with a single
power law plus Galactic column density and found a poor fit, as
expected. The residuals, shown in Figure~2b, present large-amplitude
variations in the Fe~K line region, 5--7~keV, and a trend of rising
continuum above these energies.

The XIS data were then fitted with the same model used for the
\xmm\ data, i.e., two power laws, where the second power law
parameterizes the rising continuum at higher energies. The fit yields
photon indices for the two power laws $\Gamma_1$=2.3 and
$\Gamma_2$=1.4, respectively. Again, there is a modest excess column
density which, if intrinsic, is around $\Delta N_H \sim 3 \times
10^{20}$ \nh, very similar to what measured by EPIC.  The best-fit
parameters are reported in Table~2b. The observed fluxes are
F$_{0.5-2~keV} \sim 1.8 \times 10^{-11}$ and F$_{2-10~keV} \sim 3.1
\times 10^{-11}$ \flux. Thus the flux during the \suzaku\ observation
is consistent with the flux measured by \xmm\ 2 years earlier.

Joint fits were also performed for the EPIC and XIS data, using only
the first epoch for the \xmm\ observations for simplicity. The results
are shown in Table~2c. This fit yielded improved constraints in the
Fe~K line region (\S~4.3).

From the fits to the medium X-ray energies, then, it appears that a
slight excess column density is present in \3c\ over the Galactic
value, in agreement with earlier findings (Grandi et
al. 1999). However, we shall see later (\S~4.4) that with a more
detailed parameterization of the broad-band continuum in the
0.5--100~keV energy range the slight absorption excess disappears. In
its place, a high-energy cutoff at 160~keV for the power law continuum
is requested to model the primary continuum (\S~4.4 and Figure~5).

\subsection{The Fe~K emission region} 


As apparent in Figure~2a-b a prominent emission line is present at
6~keV in both the EPIC and XIS datasets, coincident with the
redshifted \feka\ line discovered with \asca\ (Eracleous et
al. 1996). A more detailed view of the Fe~K region is presented in
Figure~3a, which shows a blow-up of the 4--9~keV region in the EPIC pn
from the A (black) and B (red) datasets and from the XIS03 data
(green). In all cases the shape of the spectrum consists of a narrow
line at an observed energy of 6~keV, easily identified with the
\feka\ emission line at the rest energy of 6.4~keV; and a weak, broad
bump as a baseline between 5.5--7.5~keV.  Thus we added two Gaussian
lines to the baseline continuum model in Table~2 to account for the
narrow emission line at 6~keV and the broad bump.  The results are
listed in Table~2 for the various datasets.  

In the case of EPIC (Table~2a), the addition of the second broader
Gaussian is significant at 99.7\% ($\Delta\chi^2$=15 for 3 additional
dofs). Its centroid energy and width are 6.6~keV and 0.5~keV,
respectively. For the XIS data (Table~2b), the addition of the broader
Gaussian improves the fit to the data by $\Delta\chi^2$=12, with P$_F
\sim$ 99.4\%. 

As apparent in Figure~3a, despite the 26 months separating the two
observations, the line profiles are very similar in both the \xmm\ and
\suzaku\ datasets.  Thus, we performed joint fits to XIS and EPIC.
The two datasets were fitted with a double power law plus two Gaussian
lines; the best-fit parameters are reported in Table~2c and the
residuals are shown in Figure~3b. The cross-normalization factor
between the XIS and PIN data was set at 1.09, as recommended for XIS
nominal observations processed before July
2008\footnote[3]{ftp://legacy.gsfc.nasa.gov/suzaku/doc/xrt/suzakumemo-2007-11.pdf}.
Table~2c shows that the fit to the joint XIS and EPIC dataset is good
($\chi^2=$3062/2728). The \feka\ line at 6.4~keV is narrow, with width
$\sigma_G=0.07^{+0.08}_{-0.04}$~keV and unresolved at \gtsima 96\%
confidence; its FWHM=8,800 km/s is consistent within a factor two with
the width of the Balmer emission lines (see \S~4.5).  The line at
6.6~keV has $\sigma_G=0.40 \pm 0.30$~keV or FWHM=43,000~km/s,
sugesting an origin from hot gas closer to the central engine.
Specifically, the line energy is consistent with emission from FeXXV
K$\alpha$.

In Table~2c both the narrow and the broad lines are highly
significant, with $\Delta\chi^2$=225 and 27, respectively,
corresponding to P$_F >$ 99.999\%.  In particular, the detection of
the broad line is robust and independent of the parameterization of
the underlying continuum. For example, fitting the 2--10~keV continuum
of \3c\ with an absorbed power law plus a narrow Gaussian at 6~keV
yields a null intrinsic column density, but still requires a broad
Gaussian line. Both narrow and broad lines are statistically required
in all datasets, as summarized in Table~3, with their significance
increasing to $>$ 99.999\% in the joint fits compared to 99.5\% for
the two datasets separately. 

We note that an intrinsically curved continuum with a cutoff around
3--4~keV, such as that expected from intrinsic cold absorption, is
highly unlikely in this type-1 AGN. Moreover, there is no evidence
from the longer wavelengths and no related absorption features were
detected in the XIS in the Fe~K region (Fig.~3). This physical
argument supports a line interpretation of the bump in 5--7~keV. 

In summary, the \xmm\ and \suzaku\ observations of \3c\ confirm the
presence of a narrow \feka\ emission line at 6.4~keV. Both datasets
independently show, for the first time, that a broad bump dominates
the 5--7~keV emission. The broad line has a rest-frame center energy
of 6.6~keV, consistent with the K$\alpha$ transition from He-like Fe
(FeXXV), thus suggesting the presence of highly ionized gas in the
inner regions of \3c.

\subsection{The spectrum at energies $>$ 10~keV} 

The PIN spectrum from 15--70~keV can be fitted well with a single power
law model with photon index $\Gamma=1.8 \pm 0.2$. The observed flux
extrapolated to the 15--100~keV band is $9.3 \times 10^{-11}$ \flux.  

A fit to the 8-channel BAT spectrum in the 15--100~keV band with a
single power law model yields an acceptable fit ($\chi^2=1.5/5$) with
photon index $\Gamma_{BAT}=2.0 \pm 0.1$ and a 15--100~keV flux of 8.2
$\times 10^{-11}$ \flux, within 12\% of the PIN flux. Because of this
and the slope similarity joint fits between the PIN and the BAT were
justified. This is especially important in view of the extended
sensitivity to higher energies provided by the BAT.

We performed joint fits to all the combined datasets:
EPIC+XIS+PIN+BAT, in the energy range 0.5--100~keV. Again, the XIS/PIN
cross-normalization was fixed to 1.09, while the one between
\suzaku\ and BAT was left as a free parameter as the two observations
are not simultaneous; the latter is consistent with 1 as indicated by
the similar HXD and BAT fluxes. The data were fitted at first with a
model consisting of a power law with a high-energy cutoff, plus two
Gaussians to model the \feka\ lines at 6.4 and 6.6~keV. Reflection off
a cold slab (\verb+pexrav+ in XSPEC) was added as well, as this
emission component was detected with \sax\ (Grandi et al. 1999) and is
expected due to the presence of the \feka\ line. The cutoff energy of
the power law was left as a free parameter, and its photon index was
tied to the reflection component's. The inclination angle of the disk
was fixed to 26\deg\ (\S~2); allowing it to vary in the range 25\deg
-- 35\deg, the total range derived from the radio and optical data,
did not impact significantly the results. The Fe abundance was fixed
to solar. 
The energy and width of the broad line were fixed at the values from
the fits to the EPIC and XIS data in Table~2c.

The results of the fit with this model are reported in Table~4a. We
find that there is no need in this model for an excess absorption
term: leaving the column density free to vary yields values consistent
to the Galactic value. A possible explanation is that the apparent
extra absorption column was mimicked by the subtle curvature of the
continuum due to the high-energy cutoff, underlying the risks of
narrow-band X-ray modeling. The best-fit photon index is $\Gamma=1.72
\pm 0.02$, consistent with the value usually found for radio-loud AGN
(e.g., Grandi, Malaguti, \& Fiocchi 2006). The data with the model,
and the corresponding residuals, are shown in Figure~4a-b. Figure~5
shows the confidence contours for the reflection strength, R, versus
the cutoff energy, clearly demonstrating the need for the latter and
the constraints on both.  While formally the fit is acceptable, with a
reduced $\chi^2_r$=1.18, Figure~4b shows non-unity residuals above
60~keV.

The presence of the 6.6~keV line raises the interesting possibility
that an ionized disk, and thus a continuum component from reflection
from an ionized medium, may be present in \3c, as suggested earlier by
Ballantyne et al. (2002). To test for this possibility, reflection
from an ionized layer of gas (model \verb+reflion+; Ross \& Fabian
2005) was explicitly included in the fit in lieu of the broad Gaussian
line. The final model consists of a power law with high-energy cutoff,
plus cold and ionized reflectors, plus a narrow Gaussian at 6.4~keV,
plus a \verb+kdblur+ component to model smearing due to relativistic
effects. The best-fit parameters are listed in Table~4b. Again the
value of the inclination angle has no impact on the model parameters
and was fixed at 26\deg\ in the fits. The shape of the ionized
component around 6.6~keV naturally accounts for the ``bump''
previously modeled with a broad Gaussian. We find that both the cold
and the ionized reflectors are required at high
significance. Specifically, without the \verb+reflion+ component the
fit with a cold reflector + Gaussian line gives a fit which is worse
by $\Delta\chi^2$=31.  The data and the model are shown in
Figure~4c-d. The addition of the ionized reflector improves somewhat
the residuals at the higher energies.

For the cold reflector, the albedo is $R=0.49^{+0.17}_{-0.14}$ and the
narrow component of the line at 6.4~keV has EW$_N$=40$^{+4}_{-9}$ eV
against the total continuum. For an inclination of 26\deg\ and the
observed photon index of $\Gamma=1.6$ we expect EW$_N$=75~eV for solar
abundances (George \& Fabian 1991). Considering the uncertainties on
the involved (fixed) parameters, such as e.g., abundances, we can
conclude that the narrow line is consistent with reflection off
Compton-thick matter subtending a solid angle R=0.5 to the
illuminating continuum, arising from reflection from the outer
accretion disk/BLR. The cutoff energy of the power law is well
determined at 157~keV, thanks to the addition of the BAT. The
ionization parameter $\xi$ of the warm reflector is high, $\xi \sim
2700$, suggesting that Fe is mainly in He-like form, thus consistent
with the results from the simple Gaussian parameterizations of the
\feka\ line profile (Table~2c).

The ionized disk component contributes 20\% of the total continuum
flux above 10~keV, which is dominated by the cold component as
observed in Figure~4c. From this perspective, the ionized reflection
strength is determined exclusively by the shape of the reflection bump
around 6.6~keV. The latter sets only marginal constraints on the 
inner disk radius. Assuming an emissivity profile proportional to
r$^{-3}$, the 90\% lower limit to the inner radius of the disk is
20r$_g$, for a fixed outer radius at 1000 r$_g$ and an inclination
angle of 26\deg. However, this result depends on the assumed disk
emissivity law: with a flatter emissivity law, $\propto r^{-2}$, the
limits on the inner radius are relaxed to smaller values (2--3 r$_g$).

\subsection{The Optical Spectrum}

We fit the H$\alpha$ region of the Keck ESI spectrum with the IMP
spectral fitting code. The narrow [O {\sc i}], [N {\sc ii}], [S {\sc
ii}] and H$\alpha$ lines were fit by double gaussian profiles with
component widths in the range 170--700 km s$^{-1}$. We take the
redshift of the [S {\sc ii}] lines ($z=0.05569$) as the galaxy rest
frame for the purpose of fitting the broad H$\alpha$.

The broad H$\alpha$ was fit with a diskline profile, convolved with a
1000 km/s turbulent velocity profile. The spectrum with the model is
shown in Figure~6. The disk emissivity profile was assumed to follow
$r^{-3}$, which corresponds to illumination by a central point-source
above the surface of the disk (see \S~4.4). The model which best fits
the line width (FWHM $\sim 12,000$ km s$^{-1}$) and separation and
relative strength of the blue and red peaks has an inclination of
$i=26 \arcdeg$ and inner and outer disk radii of 450 and 1100 r$_G$,
respectively. However, the observed broad H-alpha profile deviates
from the model in two respects. As was noted by Eracleous \& Halpern
(1994, and references therein), the ratio of blue/red peak flux is
larger than predicted by any diskline model. This suggests asymmetric
disk emission, such as might be produced by a spiral arm feature. It
is worth noting that the blue excess in the profile is very similar to
what was seen in another nearly contemporaneous observation, taken on
2004 September 21, by Gezari, Halpern, \& Eracleous (2007), while it
was not present in other epoch spectra shown by these authors. Second,
there is a very broad component (FWZI$\sim 4\times10^4$ km s$^{-1}$)
to the observed line, which could come from closer to the black hole,
or perhaps scattering by hot electrons.

We then investigated whether or not the H$\alpha$ profile is
consistent with \feka\ line profile observed by \xmm\ on the same
night. We used \verb+XSPEC+ to fit the 2--10~keV region of the EPIC pn
data from observation A with a power law continuum plus the best-fit
H$\alpha$ diskline model profile, only allowing the continuum and line
normalization to vary.  This produced a satisfactory fit to the EPIC
continuum and neutral \feka\ line, with $\chi^2=837$ for 791 degrees of
freedom. However, an equally good fit can be obtained with a simple
Gaussian. We conclude that at the EPIC pn resolution, it is impossible
to distinguish the double-peaked H$\alpha$ profile from a Gaussian
line of similar width. (We note that the XSPEC diskline profile does
not include turbulent broadening, but this is insignificant compared
to the instrumental broadening.)

Thus, an optical spectrum simultaneous to the first \xmm\ observation
confirms the presence of a double-peaked H$\alpha$ line with a width
FWHM $\sim$ 12,000 km/s and a profile consistent with disk emission
from 450 $<$ r$_G <$ 1100. The EPIC resolution is, however,
insufficient to distinguish between a simple Gaussian profile and the
H$\alpha$ best-fit profile for the \feka\ line at 6.4~keV.

\section{The Broad-band Spectral Energy Distribution}

In order to define the Spectral Energy Distribution (SED) of \3c\ we
compiled measurements from a variety of sources in the literature; the
data and related references are listed in Table~5. We applied
extinction corrections in the UV through IR bands using the extinction
law of Seaton (1979) and adopting only a Galactic reddening of
$E(B-V)=0.071$ (Schlegel, Finkbeiner, \& Davis 1998). We also shifted
the measurements to the rest frame of the source by applying the
appropriate K-corrections. We also list in Table~5 two representative
fluxes at 5~GHz and 3~mm. The former is a VLBI measurement which
isolates emission from the radio core, while the latter is a
measurement made with the IRAM 30m single-dish telescope (it may
include extended emission but we expect that at that frequency the
core should dominate).

We plot the resulting SED in Figure~7 with red circles. For
wavelengths shorter than 30~$\mu$m we adopted as much as possible
measurements through small apertures that isolate emission from the
active nucleus.  Cases where contamination from extended emission
(from the host galaxy) is suspected, are identified in Table~5 and
plotted as open circles in Figure~7. Our compilation of X-ray data is
not complete, but it is enough to show the range of variability. We
have included measurements since the early 1990s from {\it ROSAT},
{\it ASCA}, {\it RXTE}, and {\it BeppoSAX}, {\it INTEGRAL}, and from
our present work. More extensive compilations of historical X-ray data
can be found in Eracleous et al. (1996) and Grandi et al. (1999).
Also plotted in Figure~7 are the contemporaneous multiwavelength data
collected by Grandi et al. (1999) from radio to X-rays (green
circles), together with the EGRET sensitivity limit (arrow) and the
\fermi\ 5$\sigma$ detection curve for a 1-year survey exposure.

The SED of the nuclear emission in Figure~7 shows two maxima in the
optical-UV portion of the SED, and again at high energies above 10~keV
as implied by the hard X-ray continuum, which - as implied by our fits
to the \suzaku\ + BAT data - cuts off around 160~keV (Table~4).  As
discussed in \S~2 the jet is not expected to provide a significant
contribution to the nuclear emission, unless the latter is very dim;
we will return to this issue later. In Figure~7, the optical-to-hard
X-ray emission is generally dominated by the accretion flow and the IR
emission by the dusty torus.

There is obvious scatter in the SED in the optical and X-ray bands,
which we attribute to intrinsic variability of the source. In Figure~7
we have plotted with vertical error bars the $\pm 1\,\sigma$
variability amplitude at specific frequencies, as determined from
intensive monitoring campaigns or from compilations of archival data
(see details in Table~5). The observed variability amplitudes support
the interpretation of the scatter as the result of variability. The
X-ray measurements plotted in Figure~7 span a factor of 3.5 in the
2--10~keV luminosity. This range includes all the X-ray measurements
in the compilations of Eracleous et al. (1996) and Grandi et
al. (1999), save for the {\it OSO7} observations from the early 1970s,
which give a 2--10 keV flux that is 3 times larger than the largest
flux presented here, albeit with large error bars.

We determined upper and lower limits to the (variable) bolometric
luminosity from 5~GHz to 100 keV by integrating the upper and lower
envelopes of the SED.  We find that the bolometric luminosity is in
the range 1--$4\times 10^{45}$~erg~s$^{-1}$. Combining this luminosity
range with the black hole mass of $(5\pm1)\times 10^8~{\rm M}_\odot$
(Nelson et al. 2004), we infer an Eddington ratio in the range
0.01--0.07, in agreement with the range 0.02--0.04 obtained
by Lewis \& Eracleous (2006).

\section{Summary of Observational Results} 

\noindent{\it The Fe~K Region - } For the first time \suzaku\ and
\xmm\ independently provide evidence that the 5--7~keV emission from
\3c\ is complex. Two components are required: a narrow line with
EW$_N$=40~eV and FWHM=8,800 km/s from ``cold'' Fe at 6.4~keV; and a broad
bump centered at 6.6~keV which suggests emission from He-like Fe. The
\feka\ and the new 6.6~keV emission component are very similar in
shape in both the \xmm\ and \suzaku\ datasets taken 26 months apart.

\noindent{\it The X-ray Continuum - } The broad-band 0.5--100~keV
continuum is well described by a power law with a photon index
$\Gamma=1.6$, with a well-defined cutoff energy at 160~keV.  Two
reflectors are necessary to interpret the high-energy spectrum of \3c,
a cold one with $R=0.5$ and an ionized one, with ionization parameter
$\xi \sim 2700$. The broader line profile is consistent with emission
from regions of the disk outside 20 r$_g$ from the central black hole,
although this result depends on the assumed emissivity profile of the
disk. 

\noindent{\it The Soft X-ray Spectrum - } No features in absorption
are detected in the very high signal-to-noise ratio RGS spectrum; a
narrow ($\sigma_L\sim$5~eV) emission line due to OVII is present at
0.564~keV, indicating a possible origin in the NLR.  

\noindent{\it The broad-band SED - } Large
scatter is present in the optical-UV and at medium-hard X-rays, due to
intrinsic variability of the AGN. The Eddington ratio for \3c\ is in
the range 0.01--0.07.

\section{Discussion} 

\subsection{An Ionized Accretion Disk in \3c} 

We presented new results from the analysis of our \xmm\ and
\suzaku\ observations of the BLRG \3c, obtained at two different
epochs in 2004 and 2006, respectively. An integrated 9-months BAT
exposure complemented the dataset extending the energy coverage to
100~keV, and a Keck spectrum provided information on the optical
emission simultaneously to \xmm. From this rich suite of observations
a remarkably uniform picture of the inner regions of \3c\ is starting
to emerge.

In both the EPIC and the XIS spectra the \feka\ emission line is
narrow (by X-ray standards), with FWHM $\sim$ 8,800 km/s. Its width is
consistent within a factor of two with the width of the broad
H$\alpha$ line measured simultaneously (\S~4.5), indicating that an
origin of the \feka\ line from the Broad Line Region is likely. It has
been argued (Eracleous \& Halpern 1994) that the broad Balmer lines
originate from the accretion disk, consistent with their double-peaked
profiles. Indeed, the BLR could form the outer part of the accretion
disk (Eracleous \& Halpern 2003, and references therein). This
material would also be responsible for the observed cold reflection of
$R=0.5$. If the \feka\ line arises from the same parts of the disk
responsible for the optical Balmer lines, a simple prediction is that
its profile should be consistent with that of the observed H$\alpha$
emission line. Unfortunately, as seen in \S~4.5, the EPIC (and indeed,
XIS) resolution at 6~keV is insufficient to distinguish a double-peaked
profile from a simple Gaussian. Higher resolution coupled with good
sensitivity are required, a task for e.g., the calorimeter on the
upcoming {\it Astro-H} mission. 

Our new and most interesting result is the detection of a broad bump
centered at a rest-frame energy of 6.6~keV, which we interpret as
emission from He-like Fe. When fitted with a Gaussian it has a FWHM of
$\sim$ 43,000 km/s, indicating an origin from ionized matter
relatively close to the central black hole. Modeling of the broad-band
continuum shows that this component arises from reflection from an
ionized ($\xi \sim 2700$) medium at \gtsima 20 r$_G$ for an assumed
emissivity profile, confirming the earlier suggestion by Ballantyne et
al. (2002). We stress that the constraints on the inner radius are
model-dependent, and if a different emissivity profile is assumed the
value of the inner radius is relaxed (see \S~4.4). On the other hand,
an emissivity profile of $q=-3$ seems to be reasonable at least for
Seyferts (e.g., Cackett et al. 2009). Evidence for a broad line due to
H- or He-like Fe~K emission was also claimed in the
\suzaku\ observations of 3C~120 (Kataoka et al. 2007; see \S~7.2).

The intrinsic photon index we measure for \3c\ after accounting for
cold and warm reflection is $\Gamma$=1.6, close to the mean for
radio-loud AGN (Grandi et al. 2006; Reeves \& Turner 1999). Previous
X-ray monitoring of \3c\ established that the photon index is
anticorrelated with the flux on timescales of days or longer, i.e.,
the spectrum softens when the source brightens, as in non-jetted AGN
(Gliozzi et al. 2009, 2003). Together with the constraints on the
high-energy cutoff of the primary power law, this implies that thermal
Comptonization dominates the emission below 100~keV in \3c, supporting
the idea that the bulk of the X-ray continuum does not originate in a
jet.  However, a beamed non-thermal component, if present at all, is
expected to contribute at much higher energies, or when the
accretion-related emission is very faint (\S~7.3).

Another interesting result from the fits with the ionized disk model
is that the inner regions of the disk do not seem to contribute to the
Fe~K emission, as suggested by the shape of the broad \feka\ line at
6.6~keV, provided an emissivity profile $\propto r^{-3}$ is assumed
for the disk. This seems to point to a structure that obscures or
suppresses the emission from regions closest to the black hole. Since
this feature may be present in at least another broad-lined radio-loud
source observed with \suzaku\ (see below), let us discuss what
plausible physical scenarios could account for it. There are three
possible situations: a) a highly ionized ion torus/ADAF occupying the
inner disk, (b) obscuration by the base of a jet, or (c) lack of
illumination. We review them in turn.

\noindent{\it a) An ion torus/ADAF -- } In an ADAF, the central
regions of the accretion disk are inflated into a hot (T $\sim
10^{12}$ K), advection-dominated and thus radiatively inefficient ion
torus. Most of the observed radiation would thus come from cooler (T
$\sim 10^9$ K) electrons in the outer parts of the disk via
synchrotron (radio to IR) and inverse Compton (optical to X-rays). The
reflection features would be weaker than in standard optically thick,
geometrically thin disks because of the smaller solid angle subtended
by the reprocessor (the disk) to the illuminating source in the ion
torus solution.

Is an ADAF solution viable for \3c?  The ADAF mode of accretion
becomes important when the accretion rate decreases below a critical
value, which is a few percent of the Eddington value, 0.01--0.1
M$_{Edd}$ (assuming standard viscosity disk parameters; Narayan
2005). In \S~5 we inferred an Eddington ratio 0.01--0.07 for \3c,
which is close to the limit for an ADAF. In LINERs and other
low-luminosity AGN where radiatively-inefficient accretion flows are
claimed to be present, the Eddington ratios are several orders of
magnitude lower, $10^{-4}-10^{-6}$ (Ho 1999), placing \3c\ on the
boundary of ADAF-forming systems (Narayan 2005). 

Another observational feature of ADAF-dominated systems is a steep
optical-to-UV slope, and a relatively large X-ray-to-UV flux
ratio. For example, from his compilation of SEDs for seven
low-luminosity AGN, Ho (1999) derives an average optical-to-UV index
$\sim$ 1.8 with a range 1.0--3.1, compared to 0.5--1.0 for classical
bright AGN. The average optical-to-X-ray index, $\alpha_{ox}$, defined
between 2500~\AA\ and 2~keV, is $\langle \alpha_{ox} \rangle = 0.9$
vs. 1.2--1.4 for luminous Seyferts (Ho 1999). Among the data plotted
in Figure~7, some were collected sufficiently close in time to allow
intra-band comparison (Grandi et al. 1999; plotted as green
dots). Focusing on these datapoints, it is apparent that significant
UV emission relative to the optical and X-rays is indeed
present. After correcting for reddening, these measurements yield an
optical-to-UV slope $\sim$ 1 and $\alpha_{ox} \sim$ 1.2. Thus the
values of the optical-to-UV and optical-to-X-ray continuum in \3c\ are
closer to classical Seyferts than to low-luminosity ADAF
candidates. Moreover, $\alpha_{ox} \sim 1.2$ is completely consistent
with the value found for classical, optically-selected AGN with UV
luminosity similar to \3c\ (Steffen et al. 2006).  We conclude that
the optical-to-X-ray emission from \3c\ is due to a standard accretion
flow. The observed variability timescale of the optical-UV flux, of
the order of weeks to months (O'Brien et al. 1998 and references
therein), indicates an origin of the optical-UV light outside 100
r$_g$.

We calculated the expected emission of a ``pure'' ADAF for a black
hole of mass 5 $\times 10^8~{\rm M}_\odot$ and accretion rate 0.01
$M_\odot$/yr, and standard accretion disk parameters as in the model
of Narayan, Mahadevan, \& Quataert (1998). We find that the expected
ADAF contribution, dominated by bremmstrahlung emission peaking at
100~keV, is more than two orders of magnitude lower than the observed
emission and the predicted jet emission (see \S~7.3).
We thus conclude that, if there is an ADAF in \3c, its contribution to
the observed broad-band continuum emission is completely
negligible. The observed optical-UV emission in this case would still
be originating in the outer region of a standard accretion disk. Thus,
whether there is an ADAF or not in \3c\ is irrelevant from the point
of view of its observed radiation.


\noindent{\it b) Obscuration by the base of the jet -- } A second
possibility is that the base of the jet accounts for the obscuration
of the inner disk regions in \3c.  Some direct observational evidence
exists that this scenario is a plausible one, at least in an FRI radio
galaxy. High-resolution radio observations of M87 show that the jet is
still broad in the inner 30--100 r$_G$, with an opening angle of
$\sim$ 60\deg, with the collimation process still ongoing up to 1000
r$_G$ (Junor, Biretta, \& Livio 1999). On the other hand, there is
also evidence that the sub-pc radio jets of low- and high-power
sources exhibit proper motions indicative of similar Lorentz factors
in the range 3--10 (Giovannini et al. 2001). Thus, it is entirely
possible that the phenomenology observed in M87 on scales less than
100 r$_G$, i.e., a broad jet, is present also in FRII sources like \3c. 


On these scales, thus, the jet particles are still relatively cold,
i.e., non-relativistic. On the other hand, we also know from the
study of gamma-ray blazars that the jet becomes dissipative only after 
100 r$_G$ (Ghisellini \& Madau 1996).  In order to absorbe the radiation
from the inner disk, the jet must be opaque for Thomson scattering,
that is, $\tau_T =\sigma _T n_e r>1$.  Since we are observing the jet at
$\sim 30$\deg\ from its axis, we can assume that the absorbing region
of the jet has a radius comparable to the inner emitting region of the
disk, $r\sim 10~r_g \sim 10^{15}$ cm, assuming $M_{BH}=5\times
10^8\,\, M_{\odot}$.

The electron density $n_e$ can be estimated assuming that the jet
carries a kinetic power $L_k=\pi r^2 n_p m_p \Gamma_L ^2 c^3$, where for
simplicity we assume $n_e=n_p$, with $n_p$ the proton density. Therefore:

\begin{equation}
n_e=\frac{L_k}{\pi r^2 m_p \Gamma_L ^2 c^3}=2\times 10^
8\frac{L_{k,47}}{r_{15}^2 \Gamma_{L,2} ^2} \,\,\, {\rm cm}^{-3}
\end{equation}

where we have assumed a kinetic power $L_{k,47}$ (suitable for
powerful FRII sources, e.g, Tavecchio et al. 2000) and the small
$\Gamma_L =2$ is motivated by the assumption that the jet is still in
the acceleration phase at these short distances (Giovannini et
al. 1999).  Thus, $\tau_T \sim$ 0.13. Note, however, that this
estimate is affected by large uncertainties. For example, the jet
composition is unknown; assuming a pair-rich jet, $n_e>n_p$, would
yield only a lower limit, $\tau_T >$ 0.13. Thus, we can not exclude
that under some conditions the jet base could provide a source of
opacity via Thomson scattering to the X-ray emission from the disk. 


\noindent{\it c) Lack of illumination -- } Another plausible scenario
is that the inner regions of the disk are not illuminated
effectively. 
Lack of illumination could be achieved if, for example, the inner jet
is narrowly collimated very close to its base and channels energy and
radiation away from the disk surface. A similar scenario was recently
proposed for the broad-line radio-loud source 4C~+74.26, where
\suzaku\ observations also indicate that the X-ray emission originate
from the outer regions of the disk (Larsson et al. 2008).  In
principle this scenario could be tested by future ALMA observations at
high radio resolution mapping directly the inner jet. 

The ``aborted jet'' model of Ghisellini, Haardt, \& Matt (2004)
provides another possible framework. In this model, the disk in
radio-quiet Seyfert galaxies is illuminated not by a corona extended
over the disk but by the heat generated in collisions by blobs of
plasma moving at velocities smaller than the escape velocity in an
aborted jet. The distance of the collisions as well as the amount of
dissipation vary, such that if highly dissipative collisions occur
close to the black hole, where gravitational bending is also at play,
the disk is strongly illuminated yielding broad \feka\ lines with
larger EW, as observed in Seyferts. In this
scenario one could hypothesize that in radio-loud sources like
\3c\ due to the larger blob ejection velocities the collisions are
less efficient/frequent and they occur at higher distances from the
black hole, effectively illuminating only the regions of the disk at
some distance from it. This would yield weaker reflection features
emerging from some distance from the central black hole.

\subsection{Comparison to Other BLRGs and to Seyferts} 

Although only two other BLRGs have published \suzaku\ observations
to-date it is interesting to compare our results to those in the
literature. A 160~ks GTO observation of 3C~120 was presented by
Kataoka et al. (2007). They found that the emission at $>$ 2~keV is
dominated by the disk, namely, a power law with $\Gamma=1.7$, a cold
reflector with R=0.6--0.8, and a narrow \feka\ line with FWHM $\sim$
7,500 km/s and EW=60~eV; interestingly, a weak broad component to the
line is also required by the XIS data, which is interpreted as
emission of H- or He-like Fe.  If neutral Fe is assumed instead, the
inclination angle is too large and inconsistent with the limits from
the radio, 1\deg$<i<$14\deg.  Below 2~keV the XIS spectrum is variable
and best described by a soft $\Gamma \sim 2$ power-law emission, which
Kataoka et al. attribute to synchrotron emission from the jet. 

The radio-loud quasar 4C~+74.26 was also observed with
\suzaku\ (Larsson et al. 2008).  Similarly to \3c\ and 3C~120, a
narrow \feka\ line with EW=86 eV is present in 4C~+74.26
together with cold reflection in the range R=0.3--0.7 and a
$\Gamma=1.8$ continuum. The profile of the \feka\ line is consistent
with emission from \gtsima 50 $r_g$ from the central black hole,
although a relativistic profile is not formally ruled out.

Thus, the currently published \suzaku\ observations of BLRGs suggest a
broadly uniform picture where the continuum above 2~keV in these
sources is dominated by an accretion flow. As for the physical
conditions of the accretion disk, in at least two sources -- 3C~120
and \3c -- the evidence points toward relatively high ionization, and
in two other sources (\3c, 4C~+74.26) possibly to something obscuring
the innermost regions or suppressing their emission.  It is fair to
say that the advent of \suzaku\ has finally provided the necessary
sensitivity around the Fe~K region to crack open the mystery of their
central engine.


We can also attempt a preliminary comparison with the results for
radio-quiet AGN.  Recent, high S/N observations of radio-quiet Seyfert
galaxies with \xmm\ and \suzaku\ have revealed a significant diversity
in their intrinsic properties. In the sample of 26 Seyferts studied
with \xmm\ EPIC by Nandra et al. (2007) the fraction of sources with a
broad component to the \feka\ line is 70\%, of which only 45\% require
a relativistic profile with emission from a characteristic radius of
$15~r_G$. Fits to the Fe~K line region with a model including
reflection from the disk and a distant torus provide mean albedoes
R=0.54 for the disk and 0.45 for the torus, for a total $R \sim
1$. Taking into account the respective dispersions, the 1$\sigma$
range for Seyferts is $R=0.63 - 1.35$. The average photon index for
the Nandra et al. sample is 1.86 with a dispersion of 0.22, and again
the 1$\sigma$ range is 1.64 -- 2.08. These results are being confirmed
by the emerging \suzaku\ observations of Seyferts, which also show a
significant spread of intrinsic photon indices, reflections strenghts,
and lack of broad components to the Fe~K line in several cases (e.g.,
Turner et al. 2009; Ponti et al. 2009; Cackett et al. 2009; Reynolds
et al. 2008). In other Seyferts (e.g., MCG--5-23-16, Reeves et
al. 2007; MCG--6-30-15, Miniutti et al 2007), the \feka\ line is still
broad and has a relativistic profile. In general, however, it seems
that in Seyferts the range of Fe~K line widths and R values is larger
than previously thought -- unresolved to relativistic for the former,
negligible to $>1$ for the latter.

From this revised perspective the distinction between BLRGs and
Seyferts appears more blurred than in the \asca\ and \rxte\ era. It is
reasonable to state at this time that BLRGs may be clustered at one
end of the distribution of X-ray spectral parameters for Seyferts,
with significant overlap. This indicates that their central engines
differ from most, but not all, radio-quiet sources. Also
interestingly, it implies that there may be a continuum of accretion
properties across the two subclasses. The fundamental question still
remains open, however: why are BLRGs capable of forming powerful jets
while those Seyferts with similar accretion properties do not? Taken
at face value our results reinforce the idea that conditions other
than, or in addition to, accretion are required to form and launch a
relativistic jet, e.g., a rapidly spinning black hole.  As the
\suzaku\ archive for both radio-loud and radio-quiet sources continues
to grow it will be interesting to explore further the relative
distribution of observed X-ray parameters for the two classes.

Perhaps the cleanest distinction between BLRGs and Seyferts so far,
aside from radio jets, is at soft X-rays. In this band Seyferts often
exhibit prominent spectral features in emission or absorption -
signature of warm circumnuclear gas on pc scales - which so
far have been scarce or absent in BLRGs. Our analysis of the XIS and
RGS spectra in this paper reinforces this statement for \3c\ (e.g.,
Figure~1).  When the nuclear continuum is heavily obscured, as in the
peculiar case of the BLRG 3C~445, soft X-ray emission lines from a
photoionized nuclear medium can be detected (Sambruna, Reeves, \&
Braito 2007; Grandi et al. 2007), a testament to the presence of gas
in the inner regions of at least some sources. However, why BLRGs
display no features in {\it absorption} at soft X-rays remains a
puzzle.  The only possible exception so far is the luminous
BLRG/quasar 4C~+74.26 (Ballantyne 2005 and references therein); here
an \xmm\ spectrum shows absorption edges of OVII and OVIII indicating
photoionized gas with column densities $\approx 10^{21}$ \nh\ and
$\log \xi \sim 60$ (Ballantyne 2005), at the lower end still of the
distributions for Seyferts (e.g., George et al. 1998). 

It must be noted, however, that very few bright BLRGs have
high-quality soft X-ray spectra available at this time. Indeed, from
an observational point of view, the study of BLRGs is plagued by the
scarcity of available sources where to search for ionized
absorption. Only 4--5 ``famous'' objects have been observed so far,
because they are the only ones bright enough for the limited
sensitivity of past X-ray detectors -- with two of them, 3C~120 and
3C~111, located at low Galactic latitudes.

If confirmed by future observations of additional sources, the
weakness or lack of absorption features at soft X-rays in BLRGs would
qualify as (one of) the major observational distinctions between these
sources and their radio-quiet sisters.  From this perspective, it is
intriguing and perhaps related that BLRGs have collimated jets, while
Seyferts have winds. Is it possible that jet formation inhibits the
production of an accretion disk wind, and thus of a warm absorber, and
vice-versa?

Helpful hints are provided by stellar-mass black holes with
relativistic jets. A recent analysis of several epochs of
\chandra\ HETGS and \rxte\ observations of the Galactic microquasar
GRS~1915+105 (Neilsen \& Lee 2009) showed that the X-ray spectrum is
dominated by a broad Fe~K emission line at 6.7~keV from the inner disk
($<$255 r$_G$) during the low-hard state, and by a highly-ionized
absorption feature with a P-Cygni profile, indicating the presence of
a disk wind, during the high-soft state. Intriguingly, as the fraction
of hard X-ray flux in the jet decreases the strength of the wind
increases, as measured by the EW of the absorption feature (Neilsen \&
Lee 2009). The latter authors conclude that ``it appears that carrying
a significant amount of matter away from the accretion disk, strong
winds can suppress jet production''. The physical process underlying
this complex interaction is unknown.

Assuming that the same fundamental processes are at work in
stellar-mass and super-massive black holes (e.g., Merloni, Heinz, \&
Di Matteo 2003, and references therein), it is tempting to extend this
scenario to AGN. We already know that there is a strong link between
the disk and the jet in AGN, in terms of the jet carrying a power
similar to that released through the accretion process (Marscher et
al. 2002; Maraschi \& Tavecchio 2003; Sambruna et al. 2006). We thus
speculate that, similarly to the states of GRS~1915+105, the formation
of powerful, collimated jets in BLRGs subtracts mass and energy that
in Seyferts is deposited in the black holes surroundings by the disk
wind and manifests itself as the warm absorber in X-ray observations
of type 1 sources. If this scenario is correct, observationally one
may naively expect a continuum of X-ray properties across BLRGs and
Seyferts as a function of radio power, as hinted at by the limited
amount of \suzaku\ data available so far. This will be easy to verify
in the near future using sizable samples of both BLRGs and Seyferts
observed with \suzaku.


\subsection{The Jet Contribution in \3c} 

To investigate in more detail the contribution of the non-thermal jet
emission to the SED of \3c\ we present in Figure~7 a set of models
which represent the expected jet emission for typical parameters, as
observed in blazars.  Because of the small beaming factor
($\delta=0.6-1.7$, \S~2), only the external jet regions on pc-scales
are assumed to be contributing.  At this distance from the core, the
dominant energy production mechanisms are synchrotron (from radio to
optical-UV) and self-Compton scattering (at higher energies), as
\3c\ is too nearby for inverse Compton scattering on the CMB to be
energetically important. We used the code of Tavecchio, Maraschi, \&
Ghisellini (1998) to compute the models in Figure~7. Briefly, we
assumed a spherical emission region with radius $R=10^{17}$ cm,
magnetic field $B=0.3$ Gauss, and a power-law energy distribution of
the relativistic electrons with normalization $K=2.3 \times 10^6$
cm$^{-3}$, $\gamma_{min}=50$, and slope $n=2.7$ as observed in the
extended \chandra\ jets of FRIIs (e.g., Sambruna et al. 2004). The
beaming factor, for $\Gamma_L=2$, is $\delta=1.75$, and represents the
maximum Doppler factor for a jet inclination angle of 33\deg. Thus,
the models shown in Figure~7 represent the case of {\it amplification}
of the jet emission in \3c.

The different curves in Figure~7 correspond to different values of the
electron energy $\gamma_{max}$ (see caption). Comparing the
theoretical curves to the SED, it can be seen that during
average-to-high flux states the predicted jet emission is negligible
and the SED is dominated by the accretion-related radiation from IR to
hard X-ray wavelengths. However, during faint intensity states of
accretion the jet may become a substantial contributor in the optical
to medium-soft X-ray band. 

Provided that the accretion disk emission rolls off before reaching
the gamma-ray band, the jet will be the sole emitter at GeV energies,
where the source can be easily detected with the \fermi\ Gamma-ray
Space Telescope (blue curve in Figure~7), confirming the earlier
prediction of Grandi \& Palumbo (2007). A detection of \3c\ at GeV
energies was not possible with EGRET, whose sensitivity was
insufficient as shown in Figure~7. Signatures of a possible jet
emission at GeV will be a soft ($\Gamma_{LAT} \sim 2$) continuum,
perhaps with detectable variability on longer timescales (\gtsima
weeks -- months).

\section{Conclusions} 

\xmm\ and \suzaku\ observations of the BLRG \3c\ revealed a broad-band
X-ray spectrum dominated by emission from an accretion flow, with
thermal Comptonization mainly responsible for the production of the
2--10~keV continuum. Both cold and ionized reflection appear in the
X-ray spectrum; for an assumed $r^{-3}$ emissivity profile and
26\deg\ inclination angle, the ionized emission is consistent with
originating from a region that is no less than $\sim 20 r_G$ from the
central black hole. If true, this suggests that the innermost regions
of the disk are obscured, perhaps by the base of the jet, or/and
suffer poor illumination. The \feka\ line at 6.4~keV is narrow with
FWHM $\sim$ 8,800 km/s and it probably originates from the optical
region of the outer disk that emits the optical Balmer lines.

There is no evidence for absorption features in the RGS spectrum of
\3c, consistent with the suggestion that BLRGs lack warm absorbers
that are instead common in Seyferts. If confirmed by analysis of
larger samples, this would be one of the main observational
differences between the two AGN classes, perhaps related to mechanisms
of jet-disk wind formation. 

Analysis of the multiwavelength SED reveals large-amplitude
variability at optical-to-X-rays. We show that the SED is generally
dominated by emission from the accretion flow, except during low
accretion states, where synchrotron and self-Compton emission from the
non-thermal jet can contribute significantly to the optical-UV and
X-ray flux, respectively. If the accretion-powered continuum declines
around hundreds of keV, as measured with \suzaku\ and BAT, the jet
emission will be the only contributor at GeV energies, where it can be
easily detected with \fermi\ within a few years of operations.

\acknowledgements

This research has made use of data obtained from the High Energy
Astrophysics Science Archive Research Center (HEASARC), provided by
NASA's Goddard Space Flight Center, and of the NASA/IPAC Extragalactic
Database (NED) which is operated by the Jet Propulsion Laboratory,
California Institute of Technology, under contract with the National
Aeronautics and Space Administration. R.M.S. acknowledges support from
NASA through the \suzaku\ and \xmm\ programs. We thank Ski Antonucci
and G. Ghisellini for interesting e-discussions. 

Research by A.J.B. is supported by NSF grant AST-0548198.  We thank
Jenny Greene for assistance with the Keck observations.

Some of the data presented herein were obtained at the W.M. Keck
Observatory, which is operated as a scientific partnership among the
California Institute of Technology, the University of California and
the National Aeronautics and Space Administration. The Observatory was
made possible by the generous financial support of the W.M. Keck
Foundation.  The authors wish to recognize and acknowledge the very
significant cultural role and reverence that the summit of Mauna Kea
has always had within the indigenous Hawaiian community.  We are most
fortunate to have the opportunity to conduct observations from this
mountain.

\clearpage

\appendix
\section{The XIS serendipitous source} 

A seredipitous source was detected in the XIS field of view at
8.5\arcmin\ from \3c\ with RA(2000)=18 44 58, DEC(2000)= 79 48 33, and
a combined XIS 0+3 count rate of 0.215 counts\,s$^{-1}$. This source
was not detected in previous X-ray or optical images of the field,
suggesting it may be a very fast transient. The XIS spectrum of the
source can be characterized by a broken-powerlaw with
$\Gamma_1=2.98\pm0.13$ below 2 keV and $\Gamma_2=2.12\pm0.06$ above 2
keV, absorbed by a column density of
$N_H=1.4\pm0.2\times10^{21}$\,cm$^{-2}$. The XIS fluxes in the 2--10
and 0.5--10 keV are $3.5\times10^{-12}$\,erg\,cm$^{-2}$\,s$^{-1}$ and
$2.8\times10^{-12}$\,erg\,cm$^{-2}$\,s$^{-1}$. If we extrapolate this
model into the HXD/PIN bandpass, then the expected flux in the
15--50\,keV band is
$1.7\times10^{-12}$\,erg\,cm$^{-2}$\,s$^{-1}$. This is 3\% of the
15--50\,keV flux of
\3c\ ($5.4\times10^{-11}$\,erg\,cm$^{-2}$\,s$^{-1}$). Thus, the XIS
serendipitous source, being quite soft, is not expected to contaminate
the HXD spectrum of \3c.





\scriptsize
\begin{center}
\begin{tabular}{lllllll}
\multicolumn{7}{l}{{\bf Table 1: Journal of Pointed X-ray Observations}} \\
\multicolumn{7}{l}{   } \\ \hline
& & & & & &\\
Observatory  & Date & Start Time & Exposure & Detector & Energy Range & Count Rate \\
             &      &    (UT)  &   (ks)   &          &   (keV)      &   (c/s)   \\
& & & & & & \\ \hline
\xmm    & 2004 October 8--9 (A) & 19:49 &32.6 &  EPIC pn &  0.5--10    & 14.66 $\pm$ 0.02 \\
        &                 &       &45.4  & EPIC MOS &  0.6--10   & 4.75  $\pm$ 0.02 \\
         &                &       &69.7& RGS2    &  0.3--1.8    & 0.601 $\pm$ 0.003 \\ 
\xmm     & 2004 October 17 (B) & 00:12 &16 &  EPIC pn &  0.5--10    & 12.77 $\pm$ 0.02 \\
        &                 &       &22.6 & EPIC MOS &  0.6--10    & 4.14  $\pm$ 0.01 \\
         &                &       &49.4& RGS2    &  0.3--1.8    & 0.507 $\pm$ 0.003 \\ 
\suzaku  & 2007 December 14--16 & 03:25 &90  & XIS1    & 0.5--10      & 2.548 $\pm$ 0.005 \\ 
        &                 &       &85.4 & XIS0,3  & 0.5--10      &  2.004 $\pm$ 0.004 \\
        &                 &       &91.4 & HXD PIN &  10--30      &  0.153 $\pm$ 0.003 \\
& & & & & & \\ \hline

\end{tabular}
\end{center}

\noindent {\bf Notes:} Exposure after screening was applied to the data. The 
count rate is the net source intensity after screening and background subtraction.

\normalsize



\scriptsize
\begin{center}
\begin{tabular}{cccccccc}
\multicolumn{8}{l}{{\bf Table 2: Best-fit to the EPIC and XIS spectra$^a$}} \\
\multicolumn{8}{l}{   } \\ \hline
& & & & & & & \\
N$_H^z$ & $\Gamma_1$ & $\Gamma_2$ & E$_G$ & $\sigma_G$ & EW$_G$ & $\chi^2$/dof & F(2-10~keV) \\ 
& & & & & & & \\ \hline
\multicolumn{8}{l}{{\bf a) EPIC pn + MOS$^b$ }} \\
& & & & & & & \\ \hline
2.0$^{+0.9}_{-0.7}$ & $2.09 \pm 0.12$ & 1.20 $\pm$ 0.05 & 6.41 $\pm$ 0.04 & 0.04$^{+0.10}_{-0.04}$ & 26$^{+24}_{-12}$ & 2964/2733 & 3.9 \\ 
  &  &  &  6.56$^{+0.18}_{-0.45}$ & 0.52$^{+0.49}_{-0.19}$ & 85$^{+72}_{-27}$  & & \\ 
& & & & & & & \\ \hline
\multicolumn{8}{l}{{\bf b) XIS }} \\
& & & & & & & \\ \hline
3.3 $\pm$ 0.3 & 2.32$^{+0.06}_{-0.04}$ & 1.44 $\pm$ 0.21 & 6.43 $\pm$ 0.02 & 0.09 $\pm$ 0.03 & 56 $\pm$ 18 & 
  1012/944 & 3.1 \\
& & & 6.33$^{+0.2}_{-0.5}$ & 0.61$^{+0.39}_{-0.23}$ & 79$^{+39}_{-27}$ &  & \\ 
& & & & & & & \\ \hline
\multicolumn{8}{l}{{\bf c) XIS + EPIC (A)}} \\
& & & & & & &  \\ \hline
2.0$^{+0.7}_{-0.9}$ & 2.16 $\pm$ 0.05 & 1.34 $\pm$ 0.11 & 6.40 $\pm$ 0.02 & 0.07$^{+0.08}_{-0.04}$ & 38$^{+38}_{-14}$ & 3062/2728 & \\
& & & 6.60$^{+0.36}_{-0.12}$ & 0.40 $\pm$ 0.30 & 65$^{+23}_{-26}$ & & \\ 
& & & & & &&\\ \hline

\end{tabular}
\end{center}

\noindent {\bf Notes:} a=The model consists of two power laws plus two Gaussians. 
A column density fixed to Galactic ($3.5 \times 10^{20}$ \nh) acts on all components; 
b=Observations A and B were fitted jointly (see \S~4.2).  

\noindent {\bf Columns Explanation:} 1=Column density at redshift z=0.056, in $10^{20}$ \nh; 
2,3=Photon indices of the power law components; 4=Gaussian line energy center in the source's rest frame, 
in keV; 5=Gaussian line width, in keV; 6=Observed EW calculated with respect to the total 
observed continuum, in eV; 7=$\chi^2$ and degrees of freedom for the model; 8=Observed flux in the energy 
range 2--10~keV in $\times 10^{-11}$ \flux.

\normalsize



\scriptsize
\begin{center}
\begin{tabular}{cccc}
\multicolumn{4}{l}{{\bf Table 3: Significance of the narrow and broad Fe~K lines}} \\
\multicolumn{4}{l}{   } \\ \hline
& & & \\
& & Narrow & Broad \\
& & $\Delta\chi^2$ (P$_F$) &$\Delta\chi^2$ (P$_F$) \\ 
& & & \\ \hline
EPIC &&  89 (4 $\times 10^{-16})$ & 15 (2.7$\times 10^{-3}$) \\
XIS  &&  138 (8$\times 10^{-26})$ & 12 (5.7$\times 10^{-3})$ \\
EPIC+XIS && 225 (7$\times 10^{-45}$) &  27 (9.6$\times 10^{-5}$) \\
& &&\\ \hline

\end{tabular}
\end{center}

\noindent {\bf Notes:} $\Delta\chi^2$ is the improvement in the fit when a Gaussian is added to the continuum model. P$_F$ is the corresponding probability according to the F-test.

\normalsize



\scriptsize
\begin{center}
\begin{tabular}{cccl}
\multicolumn{4}{l}{{\bf Table 4: Broad-band spectral fits$^a$}} \\
\multicolumn{4}{l}{   } \\ \hline
Parameter && Best-fit Value & Comment \\ 
& & &  \\ \hline
\multicolumn{4}{l}{{\bf a) Power Law + Cold Reflection + 2 Gaussians} } \\ \hline 
$\Gamma$ && 1.72 $\pm$ 0.02 & \\
E$_{cutoff}$ && 161$^{+75}_{-62}$ & Cutoff energy in keV  \\
R$_{cold}$ &&  0.81 $\pm$ 0.04   & Cold reflection strength \\
E$_N$   &&  6.41 $\pm$ 0.02 & Narrow line rest-frame energy in keV \\
$\sigma_N$ &&  0.03$^{+1.30}_{-0.03}$  & Narrow line width in keV \\
EW$_N$     &&  23$^{+7}_{-5}$ & Narrow line rest-frame Equivalent Width in eV \\
E$_B$ &&  6.6   & Broad line rest-frame energy in keV, fixed \\
$\sigma_B$ &&  0.40  & Broad line width in keV, fixed \\
EW$_B$     && 55$^{+10}_{-20}$ & Broad line rest-frame Equivalent Width in eV \\
$\chi^2$/dofs &&  3010/2553 & \\ \hline 
\multicolumn{4}{l}{{\bf b) Power Law + Cold Reflection + Gaussian + Ionized Reflection + Kdblur} } \\ \hline 
$\Gamma$ && 1.59 $\pm$ 0.03 & \\
E$_{cutoff}$ && 157$^{+89}_{-47}$ & Cutoff energy in keV  \\
R$_{cold}$ && 0.49$^{+0.17}_{-0.14}$ & Cold reflection strength \\
E$_N$   && 6.41 $\pm$ 0.02 & Narrow line rest-frame energy in keV \\
$\sigma_N$ && 0.08 $\pm$ 0.03 & Narrow line width in keV, fixed \\
EW$_N$     &&  40$^{+4}_{-9}$ & Narrow line rest-frame Equivalent Width in eV \\
$\xi$ &&  2734$^{+975}_{-990}$ & Ionization parameter \\ 
r$_{in}$ &&  200 ($>$20) & Inner disk radius in r$_G$ \\
r$_{out}$ && 1000 & Outer disk radius, fixed \\
log r  && -3 & Emissivity profile, fixed \\
$\chi^2$/dofs && 2941/2552  & \\
&&& \\
F(0.5--2~keV) && 1.5 $\times 10^{-11}$  & Observed total flux in \flux \\
F(2--10~keV) &&  3.8 $\times 10^{-11}$  & Observed total flux in \flux \\
F(10--100~keV) && 8.2 $\times 10^{-11}$ & Observed total flux in \flux \\ \hline

\end{tabular}
\end{center}

\noindent {\bf Notes:} a=Joint fits to the EPIC pn, XIS, PIN, and BAT datasets 
in 0.5--100~keV. A column density fixed to Galactic ($3.5 \times 10^{20}$ \nh) acts 
on all components.

\normalsize



\scriptsize
\begin{center}
\begin{tabular}{cccc}
\multicolumn{4}{l}{{\bf Table 5: Spectral Energy Distribution of 3C~390.3}} \\
\multicolumn{4}{l}{   } \\ \hline
& & &  \\
$\log \nu_{rest}$ & $\log$L$_{rest}$ & Note & Reference \\ 
(Hz) & (\lum) & & \\ \hline 
& & & \\
9.72      & 41.11    &     &   1 \\
11.02                & 42.31               & a   &   2 \\
12.50                & 43.87               &     &   3 \\
12.72                & 43.96               &     &   3 \\
13.10                & 44.52               & a   &   3 \\
13.32                & 43.92               & a   &   4 \\
13.42                & 44.23               & a   &   5 \\
13.42                & 44.41               & a   &   3 \\
13.50                & 44.43               &     &   3 \\
13.66                & 44.71               & a   &   4 \\
13.97                & 44.04               &     &   3 \\
14.16                & 43.65               &     &   6 \\
14.16                & 43.96               &     &   3 \\
14.16                & 44.06               & a   &   7 \\
14.28                & 44.04               & a   &   7 \\
14.30                & 43.87               &     &   8 \\
14.42                & 44.08               &     &   3 \\
14.46                & 44.10               &     &   7 \\
14.55                & 43.77--43.85        & b   &   9 \\
14.66                & 43.98--44.07        & b   &   9 \\
14.76                & 43.95--44.09        & b   &   9 \\
14.79                & 43.71--43.83        & b   &   9 \\
14.79                & 43.37--43.64        & b   &  10 \\
14.79                & 43.64                  &     &  11 \\
14.86                & 43.83--43.97        & b   &   9 \\
15.14                & 44.06                  &     &  12 \\
15.16                & 43.44                  &     &  13 \\
15.18                & 43.65                  &     &  13 \\
15.23                & 43.19--43.72        & c   &  14 \\
15.23                & 43.53--43.95        & b   &  15 \\
15.25                & 43.72                  &     &  13 \\
15.32                & 44.30                  &     &  12 \\
15.34                & 44.05                  &     &  13 \\
15.36                & 43.50--43.98        & b   &  15 \\
15.37                & 43.24--43.77        & c   &  14 \\
& & &\\ \hline
\end{tabular}
\end{center}

\newpage

\scriptsize
\begin{center}
\begin{tabular}{cccc}
\multicolumn{4}{l}{{\bf Table 5: Continued}} \\
\multicolumn{4}{l}{   } \\ \hline
& & &  \\
$\log \nu_{rest}$ & $\log$L$_{rest}$ & Note & Reference \\ 
(Hz) & (\lum) & & \\ \hline 
& & & \\
17.41                & 43.83--44.10        & b   &  16 \\
17.41, 18.41         & 43.91, 44.23        & d   &  17 \\
17.18, 18.41         & 43.63, 44.00        & d   &  18 \\
17.10, 18.41         & 43.62, 44.01        & d   &  16 \\
17.10, 18.41         & 43.93, 44.16        & d   &  16 \\
17.10, 18.41         & 44.06, 44.35        & d   &  19 \\
16.88, 18.23, 19.41  & 43.65, 44.03, 44.63 & d,e &  20 \\
18.71, 19.41         & 44.31, 44.56        & d   &  21 \\
18.01, 18.88, 19.10  & 44.02, 44.28, 44.42 & d,f &  22 \\
17.10, 18.41         & 43.77, 44.29        & d   &  23 \\
18.58, 19.41         & 44.36, 44.69        & d   &  23 \\
& & &\\ \hline

\end{tabular}
\end{center}

\noindent{\bf Notes:} 
(a) Contamination of the measured flux by extended emission from the
host galaxy (or the large-scale radio jet and lobes in the case of the
3 mm measurement) is possible. 
(b) Range of luminosities indicating the $\pm 1\,\sigma$ spread due to
variability, as measured in an intensive monitoring campaign ranging from 
a few months to approximaely a year.
(c) Range of luminosities indicating the $\pm 1\,\sigma$ spread due to
variability, as measured from archival data spanning a few decades.
(d) Groups of frequencies and luminosities define power-law or
broken power-law models that describe X-ray sepctra. 
(e) Power-law model representing the {\it INTEGRAL} measurements made
over a $\sim 1$ year survey. We have assumed a photon index of 1.65,
which reporiduces the observed flux in the 20--40 and 40--100~keV
bands and also agrees with the photon index of the {\it Swift}/BAT
spectrum presented in this paper.
(f) Broken power-law fit to the {\it time-averaged RXTE} spectrum
presented by Gliozzi et al. (2003). This spectrum was the result of a
2-month monitoring campaign, in which the X-ray flux of the source
varied by a factor of 2.2.

\noindent{\bf References:}
 (1) Alef et al. (1988),
 (2) Steppe et al. (1988),
 (3) Miley et al. (1984),
 (4) Ogle et al. (2006),
 (5) Siebenmorgen et al. (2004),
 (6) Heckman et al. (1983),
 (7) Balzano \& Weedman (1981),
 (8) Madrid (2006),
 (9) Dietrich et al. (1998),
(10) Veilleux \& Zheng (1991),
(11) Bentz et al. (2006),
(12) Martin et al. (2005),
(13) Zirbel \& Baum (1998),
(14) Zheng (1996),
(15) O'Brien et al. (1998),
(16) Leighly et al. (1997),
(17) Malaguti et al. (1994),
(18) Eracleous et al. (1996),
(19) Evans et al. (2006),
(20) Bird et al. (2007),
(21) Grandi et al. (1999),
(22) Gliozzi et al. (2003),
(23) this work.

\normalsize



\begin{figure}[h]
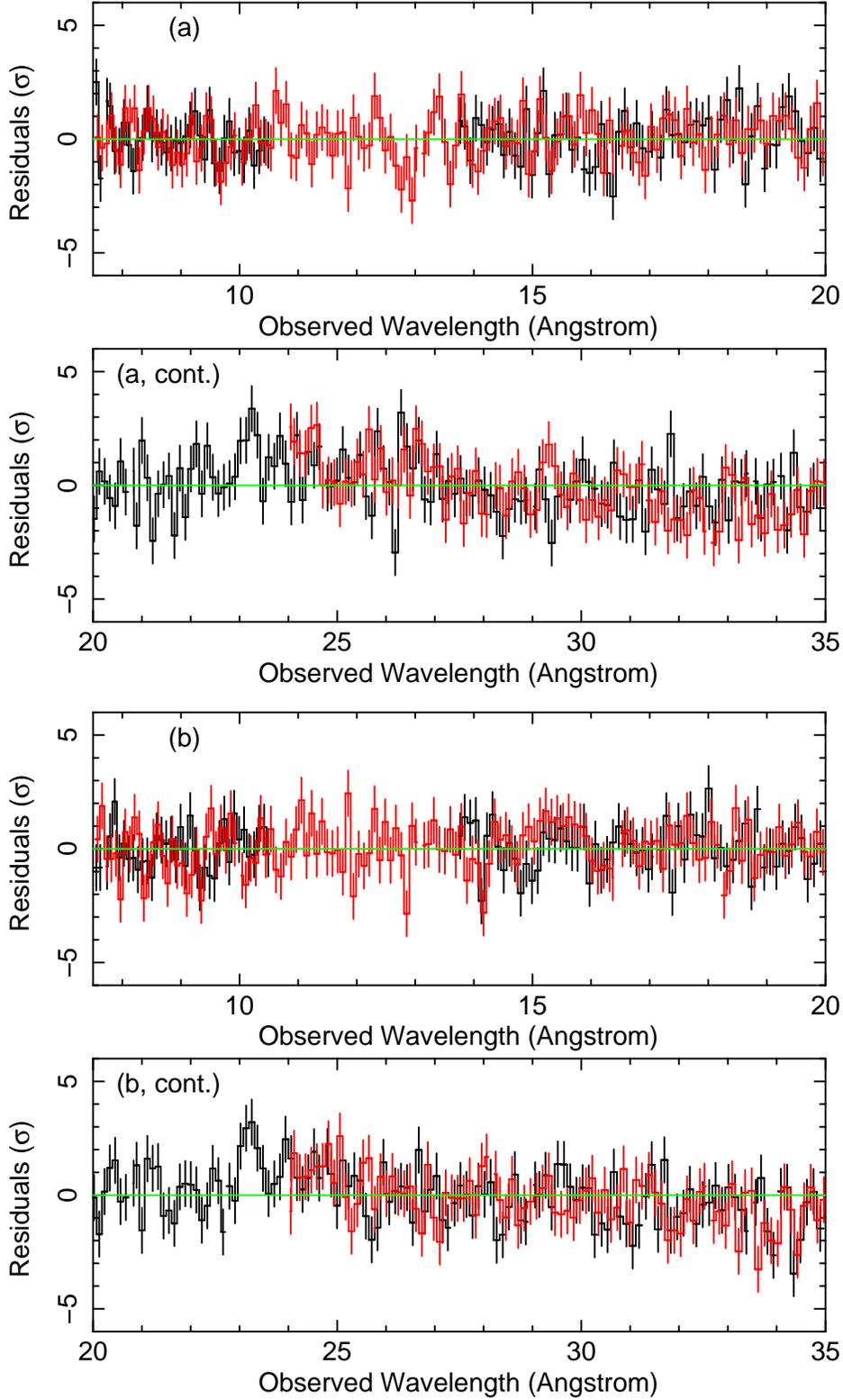

\centerline{\psfig{figure=f1a-rev.ps,height=2.in,angle=-90}}
\centerline{\psfig{figure=f1b-rev.ps,height=2.in,angle=-90}}
\vspace{0.1in}
\centerline{\psfig{figure=f1c-rev.ps,height=2.in,angle=-90}}
\centerline{\psfig{figure=f1d-rev.ps,height=2.in,angle=-90}}
\caption{\footnotesize Plots of the \xmm\ RGS spectra of \3c\ from
  observations A [(a), top two panels] and B [(b), bottom two
    panels]. The RGS1 data are plotted in black, while the RGS2 data
  are plotted in red.  Shown are the residuals of a fit with a single
  power law and a column density fixed to the Galactic value. There is
  no evidence for discrete features in absorption. A significant
  narrow ($\sigma_L\sim5$~eV) emission line is detected at
  23.2\AA\ with EW$\sim5$~eV. The line energy centroid is coincident
  with the forbidden line of OVII at 0.564~keV in the source's
  rest-frame, which may arise from the NLR on kpc scales.}
\end{figure}

\newpage


\begin{figure}[h]
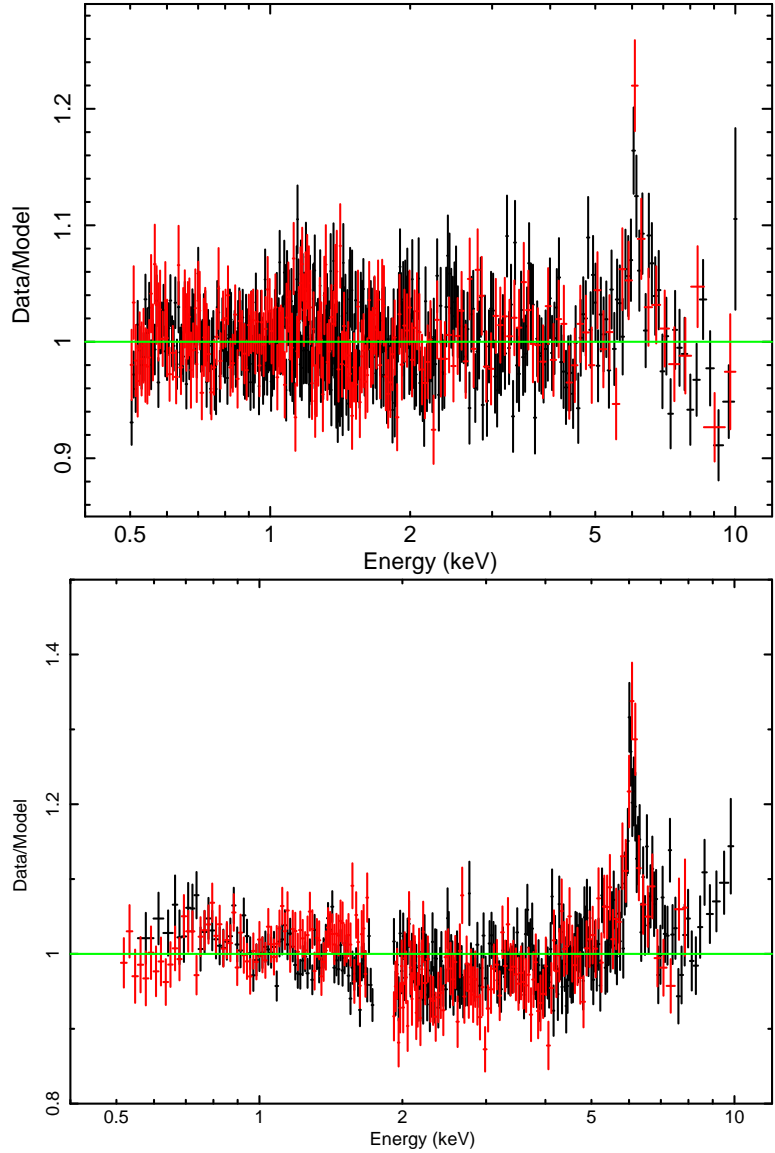

\centerline{\psfig{figure=f2a.ps,height=3.in,angle=-90}}
\centerline{\psfig{figure=f2b-rev.ps,height=3.in,angle=-90}}
\caption{\footnotesize {\it (a, Top):} Residuals of the EPIC pn
  spectra from observations A and B fitted jointly relative to a
  single power law model. {\it (b, Bottom):} Same for the \suzaku\ XIS
  data. The summed XIS03 data are shown in black, the XIS1 is in red. 
  The data gap between 1.7 and 2.0~keV is due to the detector Si
  edge, where the calibration is still not completely reliable.}
\end{figure}

\newpage


\begin{figure}[h]
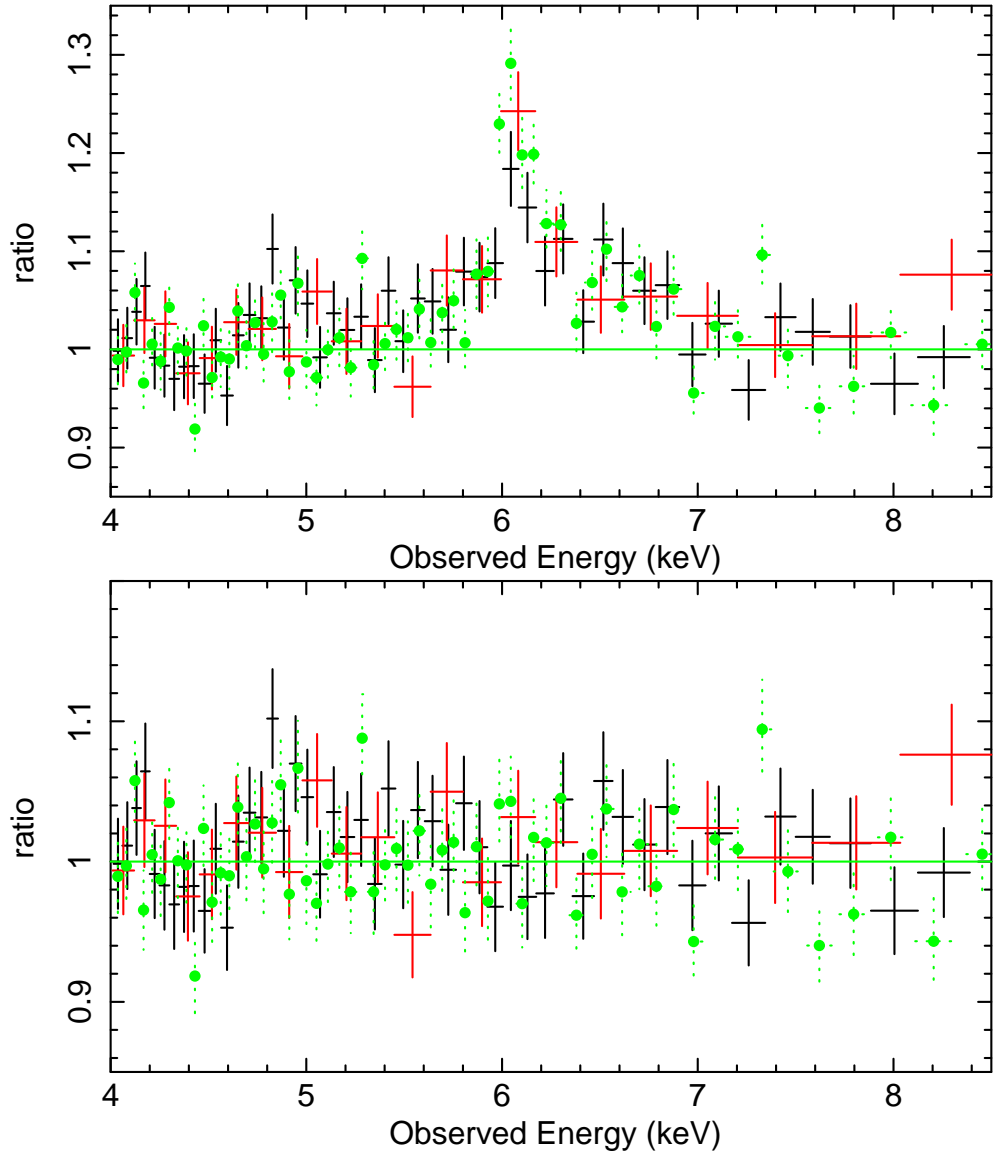

\centerline{\psfig{figure=f3a.ps,height=3.in,angle=-90}}
\centerline{\psfig{figure=f3b.ps,height=3.in,angle=-90}}
\caption{\footnotesize {\it (Top, a)} Residuals of the fits to the
  XIS03 (green dots and dotted vertical lines) and EPIC (black and
  red) observations in the energy range of the \feka\ emission line with a
  single power law model. A similar structure is seen at all epochs,
  consisting of a narrow component centered at an observed energy of
  6~keV and a broader bump between 5.8 and 7~keV. {\it (Bottom, b)}
  Same as the top panel, but fitted with a power law plus two Gaussian
  lines (see Table~2c).}
\end{figure}

\newpage


\begin{figure}[h]
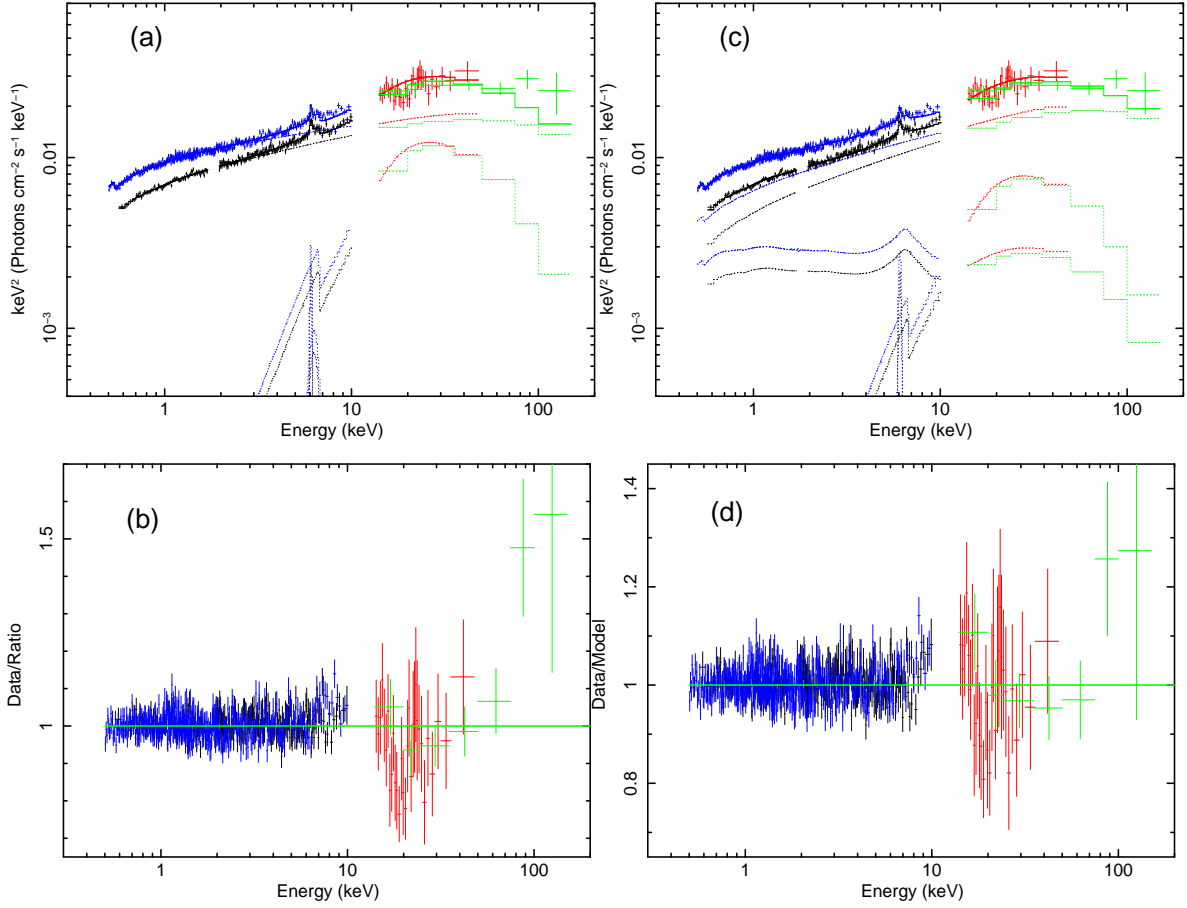

\noindent{\psfig{figure=f4a.ps,height=2.3in,angle=-90}\psfig{figure=f4c.ps,height=2.3in,angle=-90}}

\vskip.1in

\noindent{\psfig{figure=f4b.ps,height=2.3in,angle=-90}\psfig{figure=f4d.ps,height=2.3in,angle=-90}}

\caption{\footnotesize
{\it (a):} Broad-band EPIC, XIS, PIN, and BAT spectrum fitted
with the model of Table~4a, i.e., 
a power law with a cut off plus cold reflection, a narrow
Gaussian line at 6.4~keV, and a broad Gaussian centered at
6.6~keV. Only the pn observation A is shown for clarity. 
{\it (b):} Residuals of the model in (a). 
{\it (c):}
The same dataset as (a) but fitted with the model of Table~4b, i.e., 
a cutoffed power law plus cold reflection and a narrow
Gaussian line at 6.4~keV, plus ionized reflection.
{\it (d):} Residuals of the model in (c).
}
\end{figure}

\newpage


\begin{figure}[h]
\centerline{\psfig{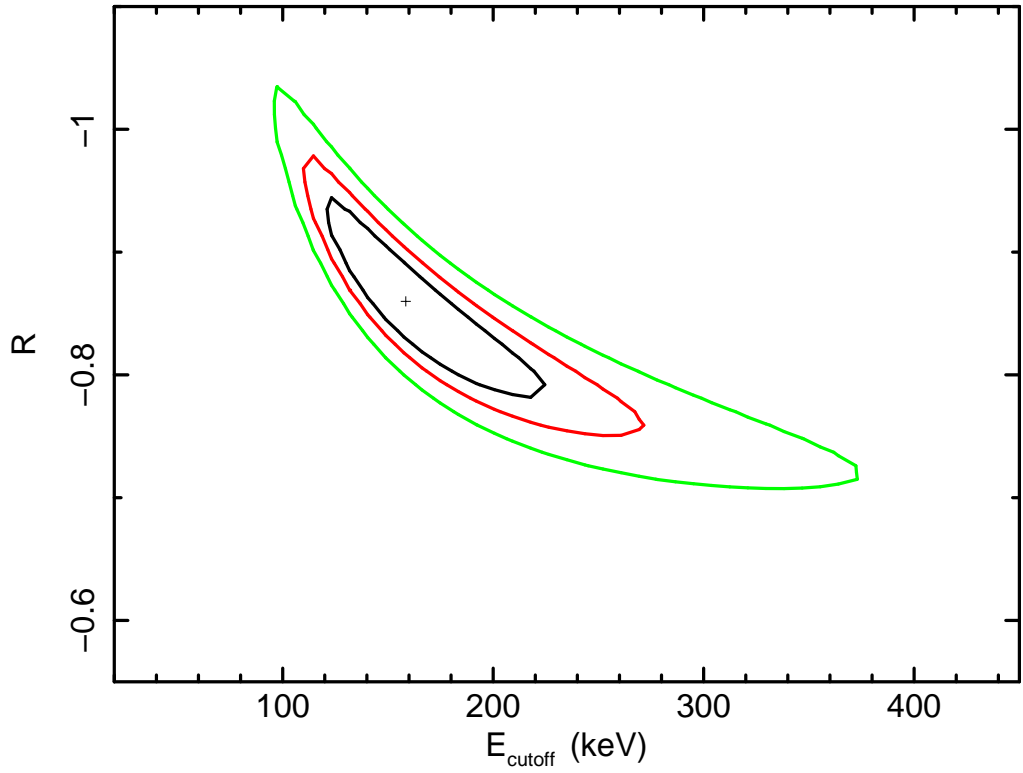}}
\caption{\footnotesize Plot of the cold reflection albedo, R, versus
  the cutoff energy of the power law, from the fits to the broad-band
  continuum with the model in Table~4a. Contours are 68\%, 90\%, and
  99\% confidence. A high-energy cutoff for the illuminating power law
  is needed and well constrained.}
\end{figure}

\newpage


\begin{figure}[h]
\centerline{\psfig{figure=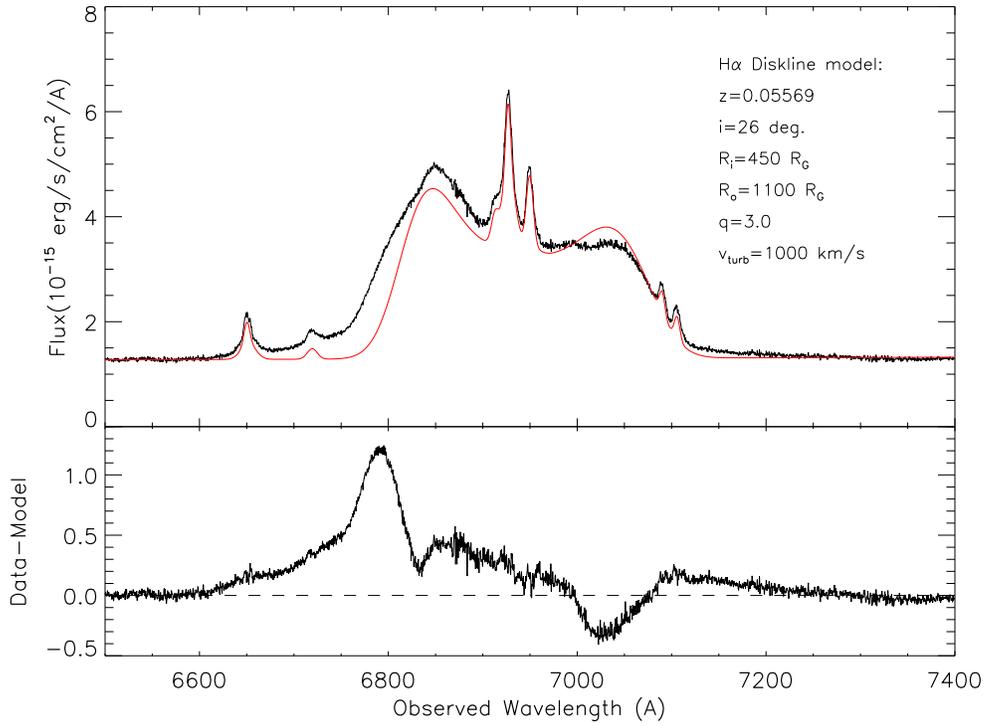,height=4.in}}
\caption{\footnotesize Zoom-in of the 
optical spectrum of \3c\ around the H$\alpha$ line. The spectrum was 
obtained at the Keck
  Observatory during the first \xmm\ exposure (observation~A in
  Table~1). The double-peaked H$\alpha$ line is fitted (top panel)
  with a diskline model with the reported parameters. The bottom panel
  shows the residuals after subtracting the line best-fit model.}
\end{figure}

\newpage


\begin{figure}[h]
\centerline{\psfig{figure=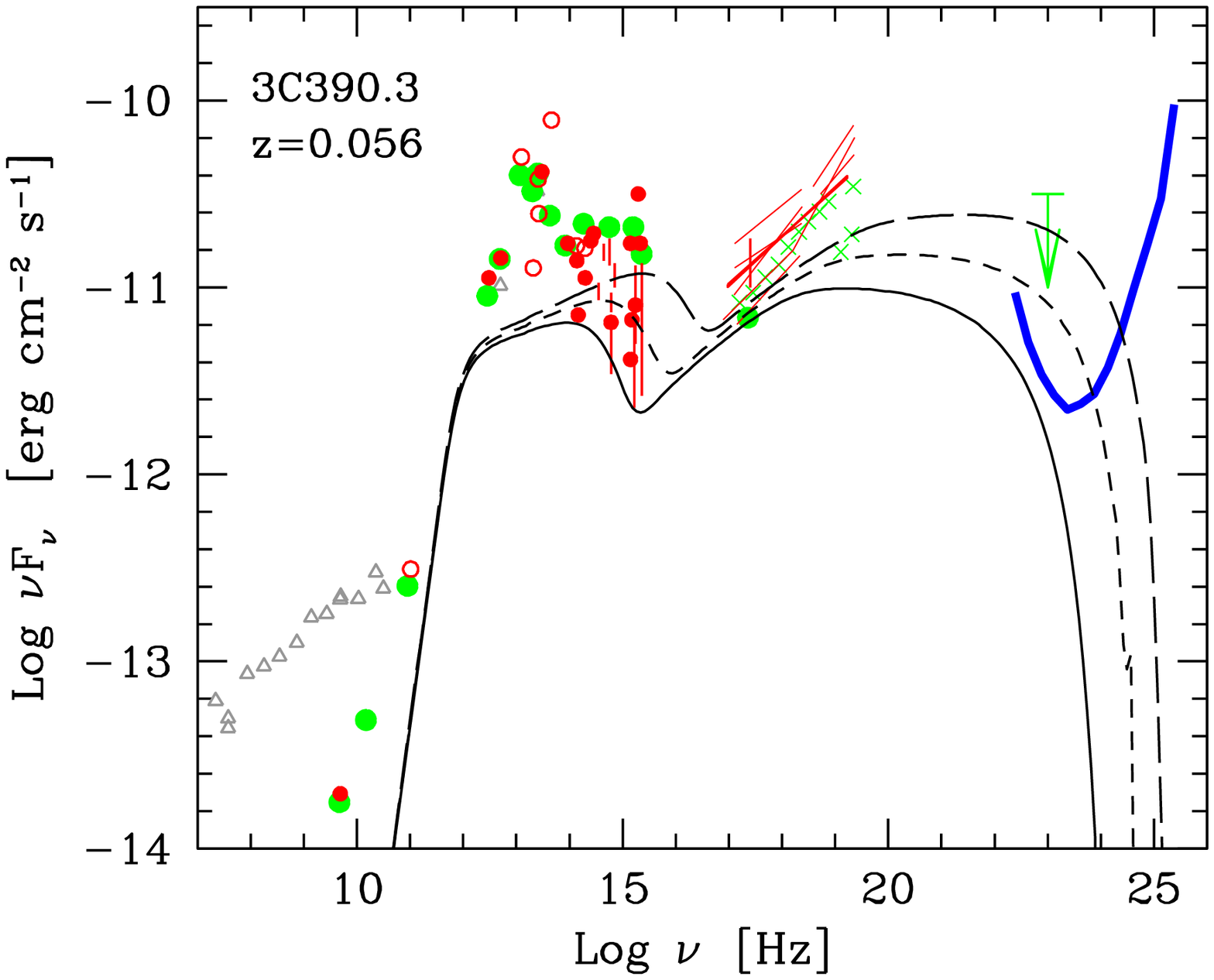,height=6.in}}
\caption{\footnotesize Spectral Energy Distribution of \3c, using the
  data in Table~5 (red circles) and the contemporaneous data from
  Grandi et al. (1999) and references therein (green circles). The
  grey triangles are from NED and represent the extended radio
  emission, included here for comparison.  
The open red dots in 
  the optical-IR indicate suspected contamination from the host
  galaxy due to the large aperture. The blue curve at gamma-rays
  is the 1 year (5$\sigma$) sensitivity threshold of the LAT
  experiment on \fermi, while the arrow indicates the previous EGRET
  upper limit. A large spread of values is visible from IR to X-rays,
  due to the source's intrinsic variability.
The continuous, dashed, and long-dashed lines are the predicted SSC
emission from the jet assuming ``typical'' blazar parameters and for
$\gamma_{max}=1.5, 3$, and $7 \times 10^4$, respectively, and beaming 
$\delta=1.75$, the maximum value for a jet inclination of
33\deg\ (\S~7.3).  The emission from \3c\ is mostly due to
accretion-related processes. However, during low accretion states the
jet can significantly contribute to the optical-to-X-ray
continuum. The jet is expected to dominate at very high energies. }

\end{figure}


\begin{references}

\reference{} Alef, W., Wu, S.Y. Preuss, E., Kellermann, K.I., \& Qiu,
Y.H. 1996, A\&A, 308, 376 

\reference{} Alef, W., Preuss, E., \& Kellerman, K. I. 1988, A\&A, 192, 53

\reference{} Ballantyne, D.R. 2007, Mod. Phys. Lett. A, 22, 2397 

\reference{} Ballantyne, D.R. 2005, MNRAS, 362, 1183 

\reference{} Ballantyne, D.R., Fabian, A. C., \& Iwasawa, K. 2004,
MNRAS, 354, 839

\reference{} Ballantyne, D.R., Ross, R.R., \& Fabian, A. C. 2002,
MNRAS, 332, L45

\reference{} Balzano, V. A. \& Weedman, D. W. 1981, ApJ, 243, 756

\reference{} Bird, A. J. et al. 2007, ApJS, 170, 175



\reference{} Blandford, R.D. 1985, in Active Galactic Nuclei,
ed. J.E.Dyson, Manchester Univ. Press: Manchester, p. 281 

\reference{} Boldt, E. 1987, PhR, 146, 215 


\reference{} Cackett, E.M., et al. 2009, ApJL, in press, arXiv:0901.3142 



\reference{} Celotti, A. \& Fabian, A.C. 1993, MNRAS, 264, 228 


\reference{} Dietrich M., et al. 1998, ApJS, 115, 185



\reference{} Eracleous, M. \& Halpern, J.P. 2003, ApJ, 599, 886 

\reference{} Eracleous, M., Sambruna, R.M., \& Mushotzky, R.F. 2000,
ApJ, 537, 654

\reference{} Eracleous, M., Halpern, J.P., \& Livio, M. 1996, ApJ,
459, 89

\reference{} Eracleous, M. \& Halpern, J.P. 1994, ApJS, 90, 1 

\reference{} Evans, D.A., Worrall, D.M., Hardcastle, M.J., Kraft,
R.P., \& Birkinshaw, M. 2006, ApJ, 642, 96 



\reference{} George, I. M. et al. 1998, ApJS, 114, 73 

\reference{} George, I. M. \& Fabian, A.C. 1991, MNRAS, 249, 352

\reference{} Gezari, S., Halpern, J.P., \& Eracleous, M. 2007, ApJS 169, 167 

\reference{} Ghisellini, G. \& Madau, P. 1996, MNRAS, 280, 67

\reference{} Giovannini, G., Cotton, W. D., Feretti, L., Lara, L., \&
Venturi, T. 2001, ApJ, 552, 508 

\reference{} Gliozzi, M. et al. 2009, ApJ, submitted 

\reference{} Gliozzi, M., Sambruna, R. M., \& Eracleous, M. 2003, ApJ, 584, 176

\reference{} Grandi, P. \& Palumbo, G.G.C. 2007, ApJ, 659, 235

\reference{} Grandi, P., Malaguti, G., \& Fiocchi, M. 2006, ApJ, 642, 113


\reference {} Grandi, P., Urry, C.M., \& Maraschi, L. 2002, NewAR, 46,
221

\reference{} Grandi, P., Guainazzi, M., Haardt, F., Maraschi, L.,
Massaro, E., Matt, G., Piro, L., \& Urry, C.M. 1999, A\&A, 343, 33 




\reference{} Guainazzi, M., Fabian, A.C., Iwasawa, K., Matt, G., \&
Fiore, F. 2005, MNRAS, 356, 295

\reference{} Heckman, T. M., Lebofsky, M. J., Rieke, G. H., \& van Breugel, W. 1983, ApJ, 272, 400

\reference{} Ho, L.C. 1999, ApJ, 516, 672 


\reference{} Junor, W., Biretta, J.A., \& Livio, M. 1999, Nature, 401,
891 

\reference{} Kalberla, P.M.W. et al. 2005, A\&A, 440, 775  



\reference{} Kataoka, J. et al. 2007, PASJ, 59, 279


\reference{} Kokubun, M. et al. 2007, PASJ, 59, 53 



\reference{} Larsson, J., Fabian, A.C., Ballantyne, D.R., \& Miniutti,
G. 2008, MNRAS, 388, 1037 

\reference{} Leighly, K.M. \& O'Brien, P.T. 1997, ApJ, 481, 15  

\reference{} Leighly, K. M. et al. 1997, ApJ, 483, 767


\reference{} Lewis, K.T. \& Eracleous, M. 2006, ApJ, 642, 711



\reference{} Madrid, J. P. et al. 2006, ApJS, 164, 307


\reference{} Malaguti, G., Bassani, L., \& Caroli, E. 1994, ApJS, 94, 517

\reference{} Maraschi, L. \& Tavecchio, F. 2003, ApJ, 593, 667 


\reference{} Martin, D. C., et al. 2005, ApJ, 619, L1


\reference{} Marscher, A.P., Jorstad, S.G., Gomez, J.-L., Aller, M., Terasranta,
H., Lister, M., \& Stirling, A. 2002, Nature 417, 625 


\reference{} Merloni, A., Heinz, S., \& Di Matteo, T. 2003, MNRAS,
345, 1057 

\reference{} Miley, G., Neugebauer, G., Soifer, B. T., Clegg, P. E., Harris, S.,
Rowan-Robinson, M., \& You ng, E. 1984, ApJ, 278, L79



\reference{} Narayan, R. 2005, Ap\&SS, 300, 177

\reference{} Narayan, R., Mahadevan, R., \& Quataert, E. 1998, in
Theory of Black Hole Accretion Disks, edited by Marek A. Abramowicz,
Gunnlaugur Bjornsson, and James E. Pringle, Cambridge University
Press, 1998, p.148

\reference{} Nandra, K., O'Neill, P.M., George, I.M., \& Reeves,
J.N. 2007, MNRAS, 382, 194 


\reference{} Neilsen, J. \& Lee, J.C. 2009, Nature, in press, arXiv:0903.4173v1

\reference{} Nelson, C.H., Green, R.F., Bower, G., Gebhardt, K., \&
Weistrop, D. 2004, ApJ, 615, 654 


\reference {} Pollock A.M.T. 2008, "Status of the RGS Calibration",  
http://xmm2.esac.esa.int/docs/documents/CAL-TN-0030-5-1.pdf


\reference{} Ogle, P., Wysong, D., \& Antonucci, R. R. J. 2006, ApJ, 647, 161

\reference{} Ogle, P. et al. 2000, ApJ, 618, 139 


\reference{} O'Brien, P. et al. 1998, ApJ, 509, 163

\reference{} Pearson, T.J. \& Readhead, A.C.S. 1988, ApJ, 328, 114 

\reference{} Ponti, G. et al. 2009, MNRAS, in press, arXiv:0901.1882



\reference{} Rees, M.J., Begelman, M.C., Blandford, R.D., \& Phinney,
E.S. 1982, Nature, 295, 17 

\reference{} Reeves, J.N. \& Turner, M.J.L. 2000, MNRAS, 316, 234 

\reference{} Reynolds, C.S., Nowak, M. A., Markoff, S., Tueller, J., 
Wilms, J., \& Young, A. J. 2008, ApJ, in press (arXiv0810:2543) 

\reference{} Ross, R.R. \& Fabian, A.C. 2005, MNRAS, 358, 211 




\reference{} Sambruna, R.M., Reeves, J.N., \& Braito, V. 2007, ApJ,
665, 1030  

\reference{} Sambruna, R.M., Gliozzi, M., Tavecchio, F., Maraschi, L., 
Foschini, L., ApJ, 652, 456

\reference{} Sambruna, R.M., et al. 2004, ApJ, 608, 698 

\reference{} Sambruna, R.M., Eracleous, M., \& Mushotzky, R.F. 2002,
NewAR, 46, 215

\reference{} Sambruna, R.M., Eracleous, M., \& Mushotzky, R. 1999, ApJ,
526, 60 



\reference{} Schlegel, D. J. Finkbeiner, D. P. \& Davis, M. 1998, ApJ, 500, 525

\reference{} Seaton, M. J. 1979, MNRAS, 187, 83P

\reference{} Sheinis, A.~I., Bolte, M., Epps, H.~W., Kibrick, R.~I.,
Miller, J.~S., Radovan, M.~V., Bigelow, B.~C., \& Sutin, B.~M. 2002,
PASP, 114, 851

\reference{} Siebenmorgen, R., Freudling, W., Krugel, E., \& Haas, M. 2004, A\&A, 421, 129

\reference{} Spergel, D. N., et al. 2003, ApJS, 148, 175

\reference{} Steffen, A.T., Strateva, I., Brandt, W. N., Alexander,
D. M., Koekemoer, A. M., Lehmer, B. D., Schneider, D. P., \& Vignali,
C. 2006, AJ, 131, 2826 


\reference{} Tavecchio, F. et al. 2000, ApJ, 543, 535 

\reference{} Tavecchio, F., Maraschi, L., \& Ghisellini, G. 1998, ApJ,
509, 608 

\reference{} Tueller, J. et al. 2008, ApJ, 681, 113

\reference{} Turner, T.J., Miller, L., Kraemer, S.B., Reeves, J.N., \& 
Pounds, K.A. 2009, ApJ, in press, arXiv:0903.4347 




\reference{} Veilleux, S. \& Zheng W. 1991, ApJ, 377, 89

\reference{} Zdziarski, A.A. \& Grandi, P. 2001, ApJ, 551, 186 

\reference{} Zheng, W. 1996, AJ, 111, 1498

\reference{} Zirbel, E. \& Baum, S. 1998, ApJS, 114, 177


\end{references}
\end{document}